\documentclass{article}
\usepackage{arxiv}
\usepackage{graphicx,color,amsmath,amsfonts,amscd,amssymb,pstricks,tikz}
\usepackage{dcolumn}
\usepackage{bm}
\usepackage{longtable}
\usepackage{amsmath}
\usepackage{mathrsfs}
\usepackage{amsfonts}
\usepackage{textcomp}
\usepackage{esvect}
\usepackage{float}
\usepackage[british]{babel}
\usepackage{environ}
\usepackage{xifthen}
\usepackage{xargs}			
\usepackage{hyperref}
\usepackage{physics}
\usepackage{multirow,here}
\usepackage{siunitx}
\usepackage{amssymb}
\usepackage{booktabs}
\usepackage{csquotes}
\usepackage{cite}
\usepackage{hyperref}

\title{Chaos indicators for non-linear dynamics in circular particle accelerators\thanks{Work funded by the HL-LHC project.}}
\author{C. E. Montanari\\
University of Manchester, Department of physics and astronomy, Oxford Rd, Manchester M13 9PL, United Kingdom \\
and \\
Beams Department, CERN, Esplanade des Particules 1, 1211 Meyrin, Switzerland \\
\And
R. B. Appleby\\
University of Manchester, Department of physics and astronomy, Oxford Rd, Manchester M13 9PL, United Kingdom \\
\And
A. Bazzani\\
Dipartimento di Fisica e Astronomia, Universit\`a di Bologna, via Irnerio 46, 40126 Bologna, Italy \\
and \\
INFN sezione di Bologna, via Berti Pichat 6/2, 40127 Bologna, Italy \\
\And
A. Fornara\\
University of Manchester, Department of physics and astronomy, Oxford Rd, Manchester M13 9PL, United Kingdom \\
and \\
Beams Department, CERN, Esplanade des Particules 1, 1211 Meyrin, Switzerland \\
\And
M. Giovannozzi\thanks{Corresponding author: massimo.giovannozzi@cern.ch}\\
Beams Department, CERN, Esplanade des Particules 1, 1211 Meyrin, Switzerland \\
\And
S. Redaelli\\
Beams Department, CERN, Esplanade des Particules 1, 1211 Meyrin, Switzerland \\
\And
G. Sterbini\\
Beams Department, CERN, Esplanade des Particules 1, 1211 Meyrin, Switzerland \\
\And
G. Turchetti\\
Dipartimento di Fisica e Astronomia, Universit\`a di Bologna, via Irnerio 46, 40126 Bologna, Italy}
\begin{document}
\maketitle

\begin{abstract}
The understanding of non-linear effects in circular storage rings and colliders based on superconducting magnets is a key issue for the luminosity the beam lifetime optimisation. A detailed analysis of the multidimensional phase space requires a large computing effort when many variants of the magnetic lattice, representing the realisation of magnetic errors or configurations for performance optimisation, have to be considered. Dynamic indicators for chaos detection have proven to be very effective in finding and distinguishing the weakly-chaotic regions of phase space where diffusion takes place and regions that remain stable over time scales in the order of multiple hours of continuous operation. This paper explores the use of advanced chaos indicators, including the Fast Lyapunov Indicator with Birkhoff weights and the Reverse Error Method, in realistic lattice models for the CERN Large Hadron Collider (LHC). Their convergence, predictive power, and potential to define a magnetic lattice quality factor linked to long-term dynamic aperture are assessed. The results demonstrate the efficiency of these indicators in identifying chaotic dynamics, offering valuable insights of these chaos indicators for optimising accelerator lattices with reduced computational cost compared to the classical approach based on CPU-demanding long-term tracking campaigns.
\end{abstract}

\section{Introduction}
Assessing the chaotic nature of the orbits of dynamical systems is a prominent research field both from the theoretical and experimental point of view. Numerous indicators have been proposed to improve chaos detection efficiency through numerical simulations. These indicators assess the system's response to initial small displacements or random perturbations along a given orbit, providing insights into chaos and long-term stability in the phase space. Recent advances in chaos theory, as well as improved computational resources, have further sparked interest in the potential applications of newer, more refined tools. Dynamic indicators have become essential tools in many fields, including celestial mechanics~\cite{PANICHI201653, Panichi2017, Lega2016fli, Guzzo2023} and accelerator physics~\cite{dynap1, PhysRevAccelBeams.23.084601, LI2021164936, montanari:ipac2022-mopost042}, where they help determine the stability of particle motion over extended timescales. For applications in circular accelerators, predicting chaotic behaviour due to multipolar magnetic field errors is especially valuable due to the possibility of studying the diffusion in the phase space and reducing the computational cost of estimating the dynamic aperture (DA).  This parameter, widely used to optimise accelerator design, is defined as the extent of the phase-space region where the orbits remain bounded for finite times and is typically computed through long-term tracking campaigns~\cite{PhysRevE.53.4067, invlog, Bazzani:2019csk}. For large and complex accelerator lattices like those of the CERN Large Hadron Collider (LHC)~\cite{Bruning:782076}, and its planned upgrade, the High-Luminosity LHC (HL-LHC)~\cite{BejarAlonso:2749422, Arduini_2016}, direct particle tracking to study long-term dynamics over physical timescales, such as \num{1e8} turns, equivalent to approximately \SI{2.5}{hours} of operation, is computationally unfeasible. As a result, there is considerable interest in novel methods and techniques capable of accurately forecasting the long-term stability and evolution of the DA, as well as enabling a greater parametric inspection due to faster DA evaluation. 

Chaos indicators have been considered for applications to problems in non-linear dynamics by various authors, and a brief list of indicators includes: the Fast Lyapunov Indicator~\cite{Lega2016fli, Guzzo2023} (FLI), its improved version, the Birkhoff Weighted Fast Lyapunov Indicator~\cite{Das_2017, Das_2018} ($\mathrm{FLI}^{\mathrm{WB}}$), the Reverse Error Method~\cite{Panichi2016,Panichi2017} (REM), and the Frequency Map Analysis~\cite{Laskar1999,Laskar2003} (FMA). Although the first two indicators have yet to be integrated into well-established optimisation frameworks commonly used in accelerator design, the third has seen rapid growth in interest, finding numerous applications within the field of accelerator physics (see, e.g., \cite{laskar1995frequency, lega1996numerical, papaphilippou1996frequency, papaphilippou1998global, Laskar1999, Papaphilippou1999, laskar2000application, PhysRevSTAB.4.124201, 1288929, Papaphilippou:PAC03-RPPG007, Laskar2003, PhysRevSTAB.6.114801, shun2009nonlinear, PhysRevSTAB.14.014001, papaphilippou2014, tydecks:ipac18-mopmf057, PhysRevAccelBeams.22.071002} for a selected list of references). This gap underscores the largely untapped potential of chaos indicators in practical optimisation tasks in the field of accelerator physics.

In a previous study~\cite{PhysRevE.107.064209}, we conducted a comprehensive performance evaluation of established and recently developed chaos indicators, focusing on their predictive power for the non-linear dynamics of a Hénon-like symplectic cubic polynomial map. The $\mathrm{FLI}^{\mathrm{WB}}$ and the REM indicators turned out to be the most effective in the classification of chaotic orbits with a limited number of iterations, so that we selected them to provide a quantitative analysis of the phase space associated with a realistic LHC lattice. We have also considered the FMA as one of the most popular methods for representing phase space of 4D symplectic maps, and we compared it with the information provided by the chaos indicators.

Yet, applying these results to real-world accelerator structures such as the HL-LHC confronts the complexities inherent in these models, necessitating the optimisation of the algorithm's computational expenses and the identification of appropriate observables to evaluate the quality factors of a magnetic lattice.

As computational tools continue to evolve, so does the potential to integrate these modern chaos indicators into accelerator physics studies. This study, therefore, represents an initial step in assessing a range of dynamic indicators for accelerator-related models, with the goal of identifying robust methods for understanding and managing chaotic behaviour in particle accelerators, and potentially opening pathways for incorporating these tools into accelerator optimisation frameworks.

In this paper, we demonstrate how the $\mathrm{FLI}^{\mathrm{WB}}$ and REM indicators can be applied to realistic HL-LHC lattices to investigate the phase-space structure near the DA. This analysis accounts for the modulational effects introduced by synchrotron dynamics on betatronic motion. We consider extensive numerical simulations of relevant cases of the HL-LHC lattice, analysing the computational cost and the predictability features of the indicators to detect the weakly-chaotic regions. For the sake of completeness, we also consider the application of other chaos indicators such as FMA to compare the results with the new indicators that we propose. 

We also consider another aspect linked with chaotic regions. In fact, the presence of extended chaotic regions is a necessary condition for diffusive behaviour of the orbits in the phase space, since there is no topological obstruction~\cite{Bazzani:262179}. We expect that a slow diffusion process occurs in the phase space~\cite{ab_fb_gt_AIP,Bazzani9948,froeschle1999weak} and that a certain fraction of the initial conditions can be lost on a long-term timescale. Therefore, a measure of the Lyapunov exponents can be related to the expected stability time $T_\mathrm{s}$. This quantity can be determined in numerical simulations by defining a control amplitude: the time when an orbit reaches such an amplitude represents the stability time. The details of the system under consideration and the particular physical application determine how the dependence of $T_\mathrm{s}$ on the arbitrary value of the control amplitude is managed.
This analysis has been performed for polynomial symplectic maps in the neighbourhood of an elliptic point~\cite{Morbidelli1995,morbidelli1995connection,CINCOTTA2022133101} suggesting the existence of a power law relation in numerical simulations. We performed a similar analysis for the complicated accelerator lattices with interesting findings.

The paper is organised as follows: Section~\ref{sec:implement} details how $\mathrm{FLI}^{\mathrm{WB}}$ and REM chaos indicators are implemented within accelerator tracking codes. Section~\ref{sec:results} presents the results of comprehensive numerical analyses, emphasising the prominent features of the chaos indicators examined. Section~\ref{sec:times} delves into the potential connection between the information provided by these indicators and the stability issues. Finally, Section~\ref{sec:conc} summarises the main findings of the investigation. Appendix~\ref{app:overview} offers a brief overview of the chaos indicators, while Appendix~\ref{app:features} discusses additional numerical simulations conducted with the HL-LHC lattice to mention the features of the chaos indicators in brief.
\section{Implementation of chaos indicators in accelerator tracking codes} \label{sec:implement}
\subsection{GPU parallel tracking of particles}
When interactions between individual particles are neglected, such as collective effects or particle-matter interactions involving secondary particle production, the single-particle tracking problem becomes a task that can profit from parallel computing~\cite{Giovannozzi:317866}. In fact, this task is suitable for parallelising initial conditions between multiple processors, since each particle's trajectory can be calculated independently, requiring only shared lattice information. As a result, tracking large numbers of particles can be efficiently parallelised on GPU architectures, which operate under the Single Instruction Multiple Data (SIMD) paradigm~\cite{DBLP:journals/corr/abs-1202-4347}. 


Since 2021, CERN has been actively developing a new symplectic tracking framework known as {\tt Xsuite}~\cite{iadarola2023xsuite, xsuite}. This innovative framework consists of a collection of {\tt Python} packages designed to enhance the capabilities originally provided by the {\tt SixTrack} code~\cite{De_Maria_2019}. {\tt Xsuite} boasts the capability of efficient parallelisation on both CPU and GPU architectures, thanks to its ability to generate optimised C code dynamically, starting from a {\tt Python} implementation. 
In particular, the {\tt Xtrack} tracking package within {\tt Xsuite} offers the ability to track particles within realistic accelerator lattices on GPU architectures. This advancement enables the tracking of a substantial number of initial conditions in considerably shorter time-frames. The application of GPUs in accelerator physics simulations has already yielded fruitful results in various studies, spanning from investigations into the Hollow Electron Lens to charged-particle tracking~studies (see, e.g.~\cite{pang2014gpu,oeftiger:hb16-mopr025,adelmann2019opal,schwinzerl:ipac21-thpab190,hermes:ipac2022-mopost045,iliakis2022enabling}). Among these diverse studies, the GPU version of {\tt Xsuite} played a pivotal role in enabling original statistical analyses of the Hollow Electron Lens~\cite{hermes:ipac2022-mopost045}, and studies of the evolution of the emittance~\cite{vanriesen-haupt:ipac2024-wepr05, fornara:ipac2024-thpc63}.

The structure of the {\tt Xsuite} codebase makes it easy to implement essential components to compute dynamic indicators within the GPU workflow. For example, the shadow-particle method~\cite{Skokos2010b, Das_2018}, a standard numerical approach used to compute the $\mathrm{FLI}^{\mathrm{WB}}$ chaos indicator, can be easily incorporated. This method involves tracking a displaced particle to monitor the evolution of its displacement while periodically normalising the particle's position. Importantly, this framework allows researchers to efficiently track many initial conditions. This capacity to handle a substantial volume of initial conditions is a key ingredient for this study, as the combination of extensive scans of the key parameters of the accelerator lattices and dynamic indicators has the potential to provide deeper insights into the phase-space characteristics of realistic accelerator lattices.

However, we stress that the complexity of the magnetic lattice of HL-LHC prevents the possibility of single-particle tracking up to the timescale of hours of operation. Chaos indicators therefore might serve as valuable tools also for predicting the stability, i.e. if the orbit reaches a prescribed amplitude value that is used as a criterion to define the motion unbounded. This means that we are stretching, more or less implicitly, the original goal of dynamical indicators, which is chaos detection, to that of assessing orbit boundedness, which, effectively, corresponds to determining the DA of the lattice. 
\subsection{Computational effort of considered chaos indicators}

In a previous paper~\cite{PhysRevE.107.064209} we studied the performance and predictability character of the different indicators to detect weakly-chaotic phase-space regions for a polynomial, time-dependent symplectic map (4D modulated H\'enon map).  
The $\mathrm{FLI}^{\mathrm{WB}}$ and REM indicators turned out to be the most effective in the classification of chaotic orbits using only a limited number of iterations. Given that the 4D H\'enon map is a good approximation of the dynamics generated by realistic accelerator lattices, we selected the same indicators to perform a quantitative analysis of the phase space of a realistic HL-LHC lattice. Note that, for the sake of completeness, we have also considered the FMA in the appendix.


In realistic applications, one has to evaluate the computational effort required by the dynamic indicators in terms of memory requirements and additional operations. Here, we summarise the key considerations for each indicator using the shadow-particle method.

Obtaining an analytical expression for the tangent map of an accelerator lattice is, in general, a highly complex task, and in the case of intricate magnetic lattices, it may even be impossible. To address this limitation, the shadow-particle method provides a straightforward numerical approach for assessing the chaotic behaviour within a non-linear magnetic lattice. However, it is known that the accuracy of this method can be influenced by the choice of the initial perturbation amplitude and the frequency of displacement resets~\cite{Skokos2010b}, an aspect that will be examined in detail in Section~\ref{sec:shadow}.

For $\mathrm{FLI}^{\mathrm{WB}}$, the primary computational effort involves tracking the orbits of two particles: the initial condition and the shadow particle. This computational load increases when examining various initial displacements concurrently. 
In contrast, both REM and FMA do not require tracking shadow particles. However, REM involves both forward and backward tracking, effectively doubling the computational effort similarly to $\mathrm{FLI}^{\mathrm{WB}}$
, FMA only needs forward tracking to compute the fundamental frequency (i.e. the betatron tune of the orbit) at different time intervals. The memory requirements for FMA may vary depending on the method used to evaluate the tune. For example, using FFT-based algorithms to calculate the fundamental frequency requires storing the complete orbit, while approaches based on the average phase advance algorithm~\cite{Bartolini:292773, Bartolini:316949} can be implemented without this requirement. We would like to mention that a notable improvement of the average phase advance method consists of combining it with Birkhoff weights~\cite{russo:ipac2021-thpab189} as this method offers super-convergence for the case of regular orbits and can be computed in a single forward tracking pass, eliminating the need to store the entire orbit history. Excessive memory requirements inevitably present challenges, making it not convenient to utilise SIMD architectures effectively, due to limitations in GPU onboard memory and the latency associated with frequent data transfers between GPU and host memory.

\subsection{Results of preliminary tracking studies with HL-LHC lattices} \label{sec:model}
To compare the performance of the dynamic indicators, we consider a realistic accelerator lattice implementation based on version 1.5 of the HL-LHC layout and optics~\cite{hllhc15update, hllhc15url} for the clockwise beam, the so-called Beam~1. The optical configuration is based on the so-called flat optics, in which the horizontal and vertical beta-functions at the interaction points are different. We conducted single-particle tracking without considering beam-beam interactions at top energy, corresponding to \SI{7.0}{TeV}. This is just a case study taken as an example, and further studies will explore both different configurations and energies, such as round optics and injection energy. 

An overview of the machine settings used for the simulation is given in Table~\ref{tab:lattice_parameters}. Our tracking simulations started at IP3, where the $\beta$ functions for the two planes are $\beta_x=$ \SI{113.22}{\meter} and $\beta_y=$ \SI{225.40}{\meter}. We implemented the lattice using the {\tt MAD-X} code~\cite{madx}, with the resulting lattice converted to track with the {\tt Xsuite} library.

\begin{table}[!hbt]
   \centering
   \caption{Main parameters of the lattice used for the tracking studies. Note that in all these studies the normalised nominal transverse emittances are the same in horizontal and vertical planes and equal \SI{2.5}{\micro m}. The rms bunch length $\sigma_\zeta$ is \SI{7.61}{cm}.}
   \begin{tabular}{lcc}
       \toprule
       \textbf{Parameter} & \textbf{Value} \\
       \midrule
Beta-functions at interaction points,           $\beta_x^\ast, \beta_y^\ast$ [\si{cm}] & $7.5, 18$ \\ 
Tunes, $Q_{x},Q_{y}$ & $62.316, 60.321$ \\
Chromaticities, $\xi_x$, $\xi_y$ & $15, 15$ \\
Current in the arc octupoles  [\si{\ampere}]& $+100$ \\
Half crossing angle, $\theta_\mathrm{c}/2\ [\si{\micro \radian}]$ & $250$\\
       \bottomrule
   \end{tabular}
\label{tab:lattice_parameters}
\end{table}

The magnetic lattice generated by {\tt MAD-X} includes 60 realisations, often referred to as seeds, of the magnetic field errors associated with the magnets in the ring. These realisations are introduced to consider the impact of the uncertainty on the measured values of the magnetic field errors on the beam dynamics. Hence, the analysis of the tracking results is usually performed by inspecting all realisations and looking at the statistical distribution of the observed features. In our case, from the set of 60 realisations, we selected two representative samples: one exhibiting the largest stable domain (referred to as ``best seed'' from now on) and one with the lowest one (the ``worst seed''). The stable domain has been estimated using a $100\times100$ uniform Cartesian grid of initial conditions of the form $(x_0, 0, y_0, 0)$, corresponding to the transverse plane $x_0-y_0$. It should be noted that linearly normalised coordinates referred to the closed orbit are used, and the unit is the nominal beam sigma, computed using the nominal value of the emittance (see Table~\ref{tab:lattice_parameters}). The largest connected set of initial conditions that survived up to \num{1e5} turns defines the stable region for the lattice configuration.

\begin{figure}[th]
    \centering
    \includegraphics[width=0.9\textwidth]{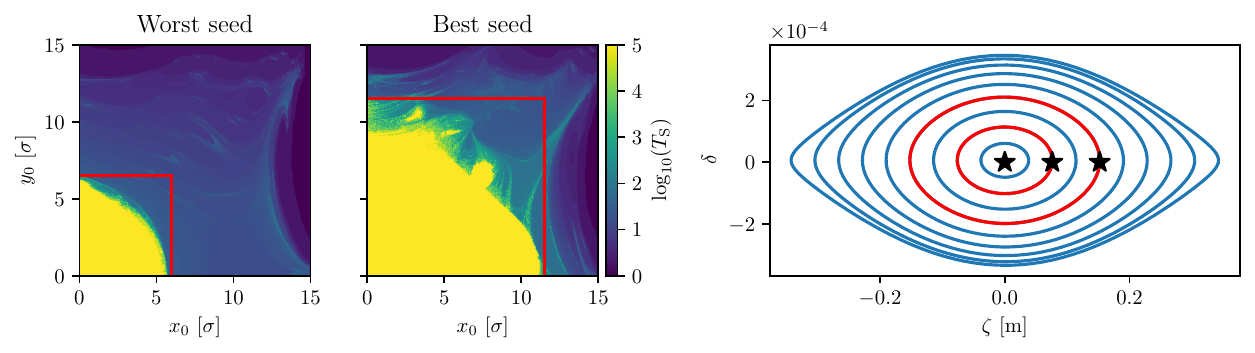}
    \caption{Left plots: representation of the tracking results of a set of initial conditions using  the HL-LHC lattice for Beam~1 at \SI{7.0}{TeV}, without beam-beam interaction. The colour encodes the logarithm of the stability time $T_\mathrm{s}$, i.e. the time over which the motion remains bounded, noting that the tracking has been performed up to \num{1e5} turns. The two plots refer to the worst and best seeds, respectively. The initial conditions are distributed over a $100\times100$ Cartesian grid in the $x_0-y_0$ plane, with a ROI highlighted in red to mark boundaries for finer sampling used in the analysis of the dynamic indicators. Right plot: orbits in the longitudinal phase space. Initial conditions with varying $\zeta$ values are tracked up to \num{1e3} turns for the best seed, illustrating the well-known pendulum-like structure with unstable fixed points around $\zeta \sim \pm \SI{0.35}{\meter}$. Black stars indicate the values of $\zeta$ at 0, 1, and 2 units of $\sigma_\zeta$, chosen for investigating the dynamic indicators in their dependence on the longitudinal dynamics.}
    \label{fig:seed_presentation}
\end{figure}

Figure~\ref{fig:seed_presentation} (left plots) presents a visualisation of the stability domain for the worst and the best seeds. The initial conditions defined on the Cartesian grid are coloured according to the logarithm of their stability time $T_\mathrm{s}$, which is the time for which the orbit remains bounded, and the yellow area represents the region inside which the orbits are stable up to the maximum number of turns used in the tracking study, i.e. \num{1e5}. The different extents of the stable region are clearly visible together with the difference in the shape of its boundary. The red boundaries represent the region of interest (ROI) that will be used in the rest of the study of chaos indicators.

Figure~\ref{fig:seed_presentation} (right plot) represents an image of the longitudinal phase space computed using the best seed configuration. Various orbits are shown, and those highlighted in red represent special values of the $\zeta$ coordinate, corresponding to $0$, $1$, and $2$ $\sigma_\zeta$, which have been used in the study of chaos indicators to probe the overall impact of longitudinal dynamics on orbit chaos.

\subsection{Considerations on the shadow-particle method} \label{sec:shadow}

The shadow-particle method requires the choice of the amplitude of the initial displacement $\epsilon_0$ of the shadow particle with respect to the reference particle, and the time interval $\tau$ between successive renormalisations of the distance between the shadow particle and the reference particle. The choice of these parameters can significantly influence the evaluation of the FLI and $\mathrm{FLI}^\textrm{WB}$ (see, e.g.~\cite{Das_2018}), since in long-term simulations on complex systems, like HL-LHC, the numerical precision of floating point operations (i.e. \num{1e{-14}}) may affect the computation of the Lyapunov indicator if the values of $\epsilon_0$ and $\tau$ are too small. In contrast, small values of these parameters are needed to compute the linearisation of the dynamics around the reference particle. In this respect, it should be mentioned that the definition of the displacement and its regular renormalisation should be performed in normalised coordinates, not in standard physical coordinates, to avoid spurious effects in the numerical calculations.

\begin{figure}
    \centering
    \includegraphics[width=0.55\textwidth]{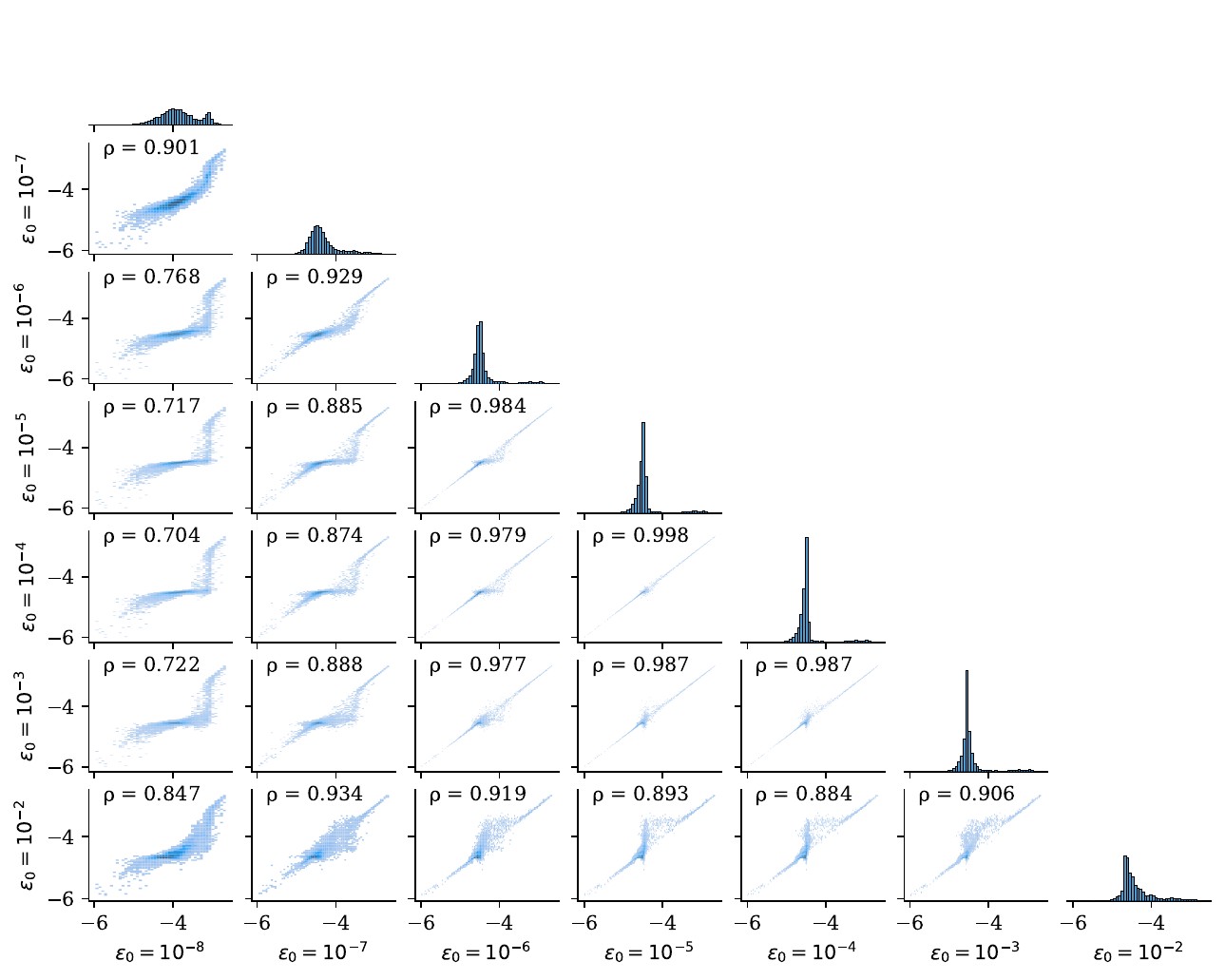}
    \includegraphics[width=0.35\textwidth]{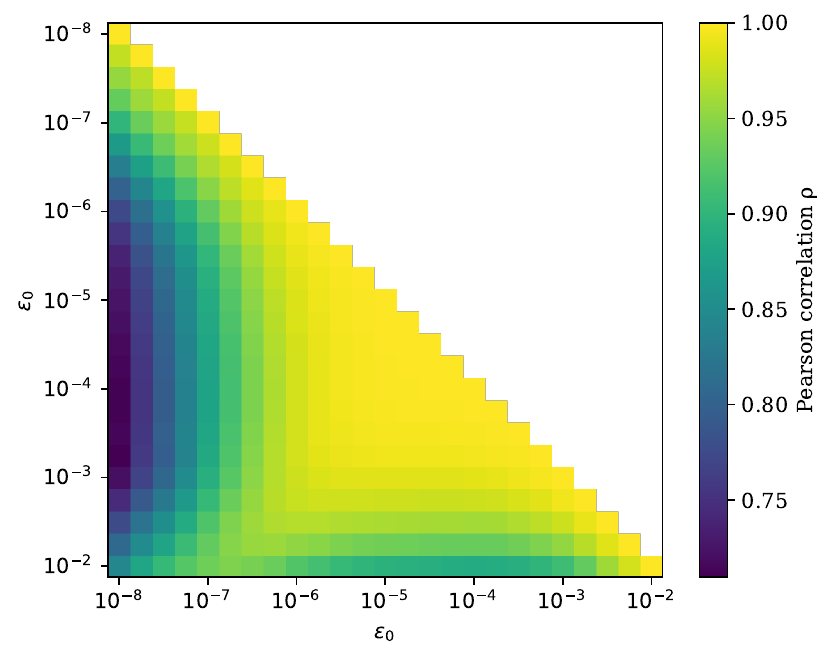}
    \includegraphics[width=0.55\textwidth]{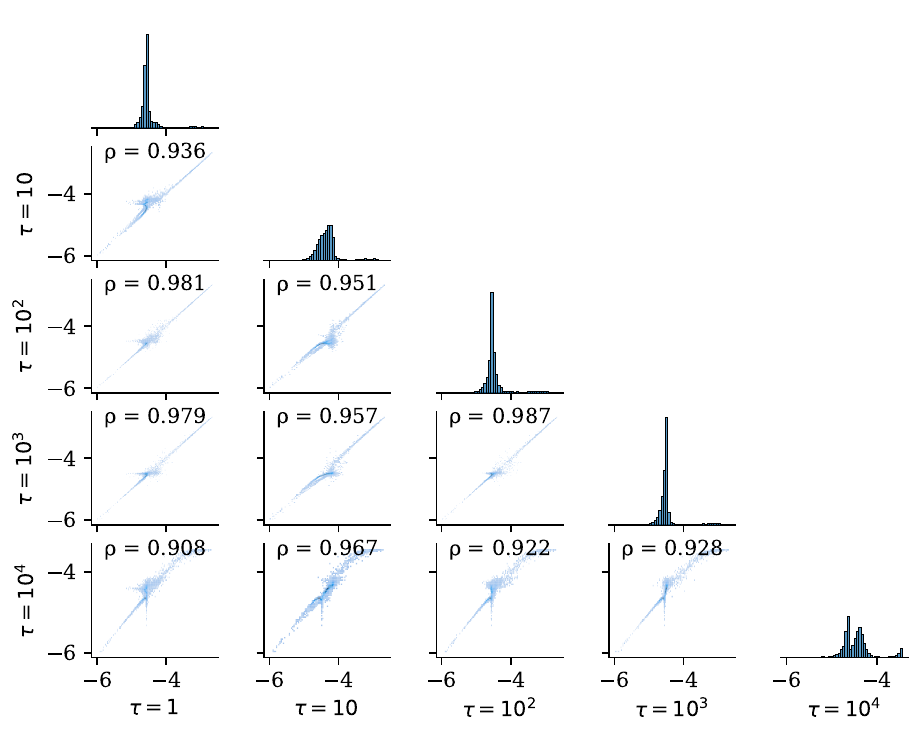}
    \includegraphics[width=0.35\textwidth]{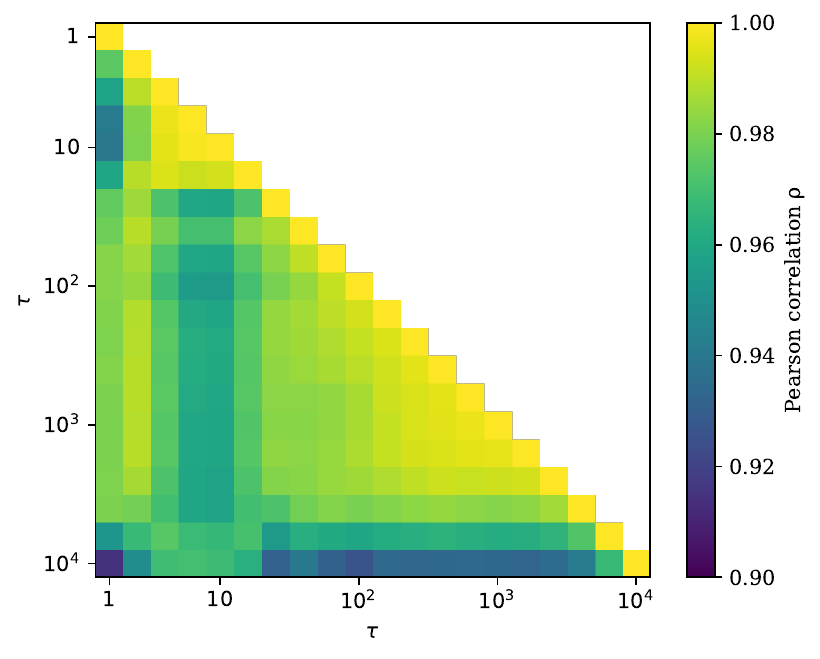}
    \caption{Pair plots and Pearson correlation analyses of $\log_{10}(\mathrm{FLI}/n)$ evaluated at \num{1e5} turns using the HL-LHC lattice with the worst seed, for various settings for initial displacement $\epsilon_0$ (top row) and the renormalisation interval $\tau$ (bottom row). The left column plots show pair plots with histograms of the indicator along the diagonal and correlation plots for different parameter pairings off-diagonally. The right column plots show a finer-grained analysis of the Pearson correlation. For $\epsilon_0$, lower values (e.g. \num{1e{-8}}) yield a wider spread of the indicator for regular initial conditions, similarly to what occurs for large values (e.g. \num{1e{-2}}). In contrast, variations in $\tau$ are less impactful, provided $\tau$ remains below \num{1e3}. For larger $\tau$ values, values of $\log_{10}(\mathrm{FLI}/n)$ exceeding $-3$ appear flattened, pointing to reduced precision at large $\tau$ settings.}
    \label{fig:epsilon_tau_fli}
\end{figure}

To address the impact of these parameters on the results of numerical simulations, a comparative study has been performed. The results are shown in Fig.~\ref{fig:epsilon_tau_fli}, where a pair plot of $\log_{10}(\mathrm{FLI}/n)$ evaluated at \num{1e5} turns is presented using the HL-LHC lattice with the worst seed for various settings of $\epsilon_0$ (top row) and $\tau$ (bottom row). To distinguish between regular and chaotic orbits, a bi-modal distribution in the histogram of the chaotic indicator is expected~\cite{PhysRevE.107.064209}. The distribution of the indicator is clearly influenced by the choice of $\epsilon_0$ and $\tau$, with the bi-modal character ultimately disappearing for certain values of the parameters.

For the remainder of this study, we will consider $\epsilon_0 = \num{1e{-4}}$, as it appears to be the best compromise between the two extremes, and the corresponding histogram distribution manifests the sharpest bi-modal distribution.

Regarding the choice of $\tau$, as long as $\tau < 10^3$, we do not observe significant variations in the evaluations of $\log_{10}(\mathrm{FLI}/n)$. Above that value, we have more significant fluctuations in the correlation plots, but the histograms of the indicators do not appear to be significantly affected. This can be understood by the fact that an excessively high value of $\tau$ can cause the shadow particle to be too far from the orbit of the reference particle, and its dynamics cannot be used to reliably determine the tangent map of the orbit of the reference particle. Hence, for the reminder of this study, we will consider $\tau = \num{1e2}$, as a good trade-off.

\section{Chaos-detection studies} \label{sec:results}


\subsection{Overview of the chaotic regions of the HL-LHC lattice}

A finer Cartesian grid made of $300 \times 300$ initial conditions has been considered in the selected ROIs, defined in Fig.~\ref{fig:seed_presentation}, to obtain an overview of the phase space that ranges from the neighbour of the closed orbit to the border of the stability region (SR) computed at \num{1e6} turns\footnote{Note that this time interval corresponds to approximately \SI{90}{ seconds} in the actual accelerator.}.

\begin{figure}
    \centering
    \includegraphics[width=0.9\linewidth]{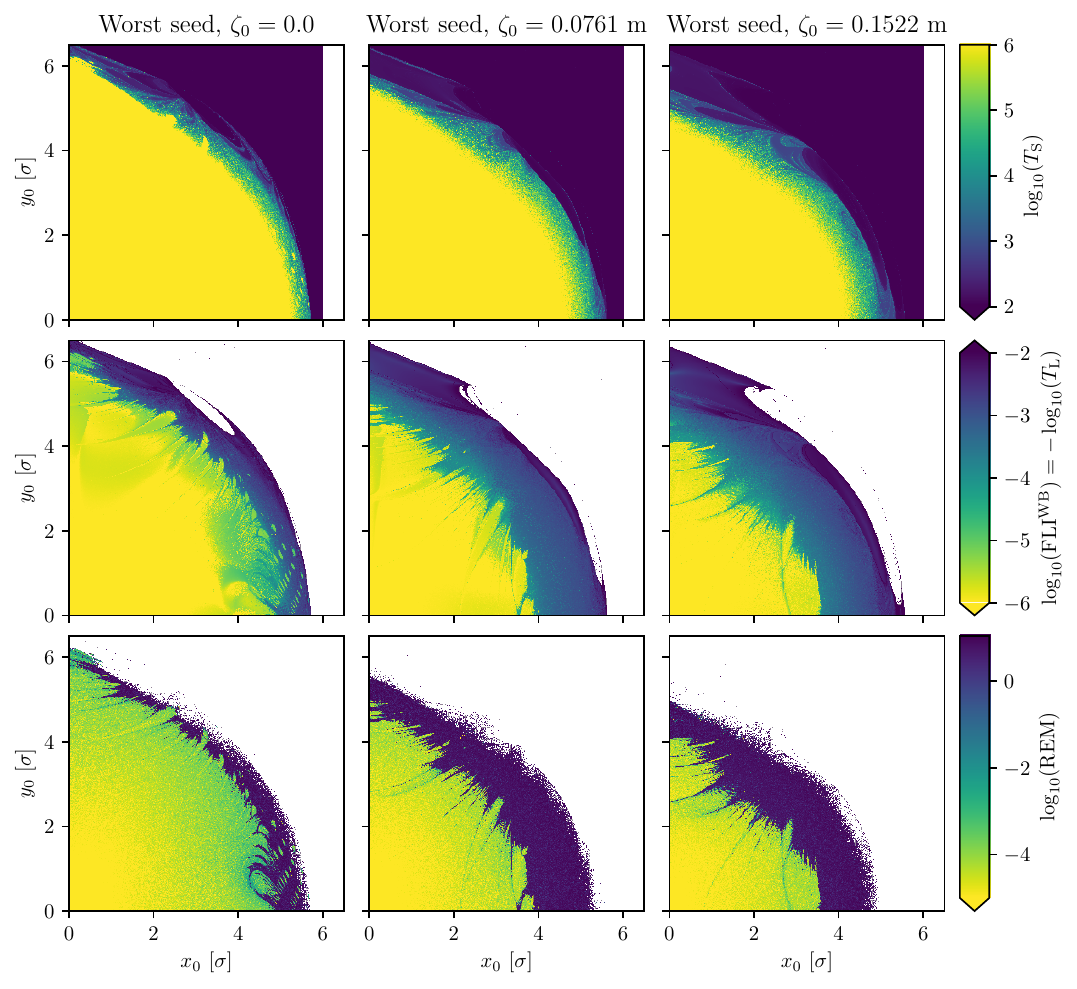}
    \caption{Overview of the phase-space structure for the HL-LHC configuration described in the text with the worst  seed and tracking performed up to \num{1e6} turns. The three columns refer to the three values of the longitudinal variable $\zeta_0$ used in the tracking studies. The top row shows the stability plots, representing the logarithm of $T_\mathrm{s}$. The middle row presents the behaviour of $\log_{10}(\mathrm{FLI}^{\mathrm{WB}}(\hat{x}))$, highlighting the chaotic regions near the boundaries of the stable area. The bottom row displays the values of $\log_{10}(\mathrm{REM})$, further identifying the transition between regular and chaotic dynamics. The reduction of the extent of the stable region for increasing values of $\zeta$ is clearly visible as well as the increase of the chaotic region. Both chaos indicators show the emergence of isolated islands of regular motion in between of chaotic regions. Note that the indicators are computed for stable initial conditions using the tracking data up to \num{1e6} turns, while for unstable initial conditions the computations are carried out using the stable part of the orbit.}
    \label{fig:full-view-worst-seed}
\end{figure}

\begin{figure}
    \centering
    \includegraphics[width=0.9\linewidth]{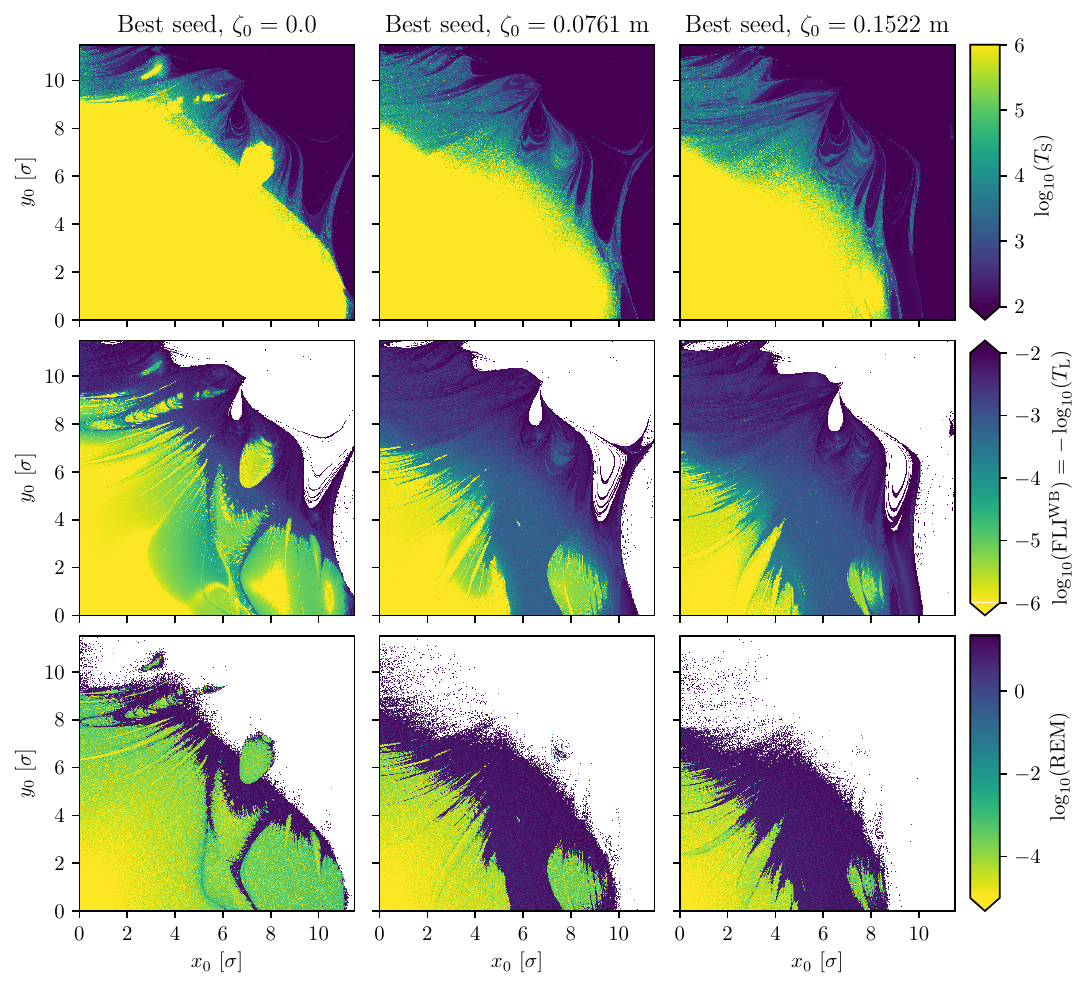}
    \caption{Overview of the phase-space structure for the HL-LHC configuration described in the text with the best seed  and tracking performed up to \num{1e6} turns. The three columns refer to the three values of the longitudinal variable $\zeta_0$ used in the tracking studies. The top row shows the stability plots, representing the logarithm of $T_\mathrm{s}$. The middle row presents the behaviour of $\log_{10}(\mathrm{FLI}^{\mathrm{WB}}(\hat{x}))$, highlighting the chaotic regions primarily at the periphery of the stable area, although some light areas are also present closer to the origin. The bottom row presents the values of $\log_{10}(\mathrm{REM})$, confirming the boundaries between regular and chaotic behaviour. The results demonstrate a more sharply defined stable region compared to the worst seed as well as wider chaotic region. Note that the indicators are computed for stable initial conditions using the tracking data up to \num{1e6} turns, while for unstable initial conditions the computations are carried out using the stable part of the orbit.}
    \label{fig:full-view-best-seed}
\end{figure}

The structure of the phase space for the HL-LHC configurations is presented in Fig.~\ref{fig:full-view-worst-seed} and Fig.~\ref{fig:full-view-best-seed}, corresponding to the worst and best seed of Fig. \ref{fig:seed_presentation}, respectively. These figures show a comprehensive view of the behaviour for each configuration. The first row of each figure shows the SR, illustrating the survival of the initial conditions up to \num{1e6} turns for different fixed values of $\zeta = \zeta_0$. In this study, we focus on three values: $\zeta_0 = 0, 1, 2\times\sigma_z$ with $\sigma_z = \SI{0.0761}{m}$. The extent of the stable region decreases with an increasing value of $\zeta_0$ for both seeds. It is worth noting that the transition between stable to unstable regions is sharper for the case of the worst seed, whereas the best seed features an almost continuous transition with structures extending to large amplitudes. 

To investigate the details of the phase-space structure, we employ chaos indicators. The second row of each figure shows the values of $\log_{10}(\mathrm{FLI}^{\mathrm{WB}}(\hat{x}))$, highlighting the emergence of chaotic regions in the transition between stable and unstable phase space. Moreover, in this scenario, the size of the chaotic regions is influenced by $\zeta_0$ and grows as $\zeta_0$ increases. It is important to note that these indicators identify chaotic regions even within what is considered the stable zone. This observation suggests that there might be initial conditions that, although chaotic, remain stable at least within the number of turns used in our tracking analyses. The exploration of the connection between stable and chaotic dynamics follows in the subsequent section. Additionally, it is noted that while the phase space geometry for the worst-case seed resembles a circle, such is not true for the best-case seed. The chaos indicator effectively points out the intricate structures, and it also exposes islands of regular motion amidst the chaotic regions. 

The third row illustrates the phase-space structure derived using the $\log_{10}(\mathrm{REM})$ indicator. In general, the REM results qualitatively align with those shown by $\mathrm{FLI}^{\mathrm{WB}}$. It should be noted that the range of values for the $\mathrm{FLI}^{\mathrm{WB}}$ indicator is quite wide, while that for REM is somewhat narrower, with a distinct clustering around extreme values. This suggests that both REM and $\mathrm{FLI}^{\mathrm{WB}}$ share a bi-modal distribution of their values. However, REM shows a more pronounced transition between the two modes, which facilitates a method for defining a threshold that distinguishes regular from chaotic orbits. 

It should be noted that the chaos indicators have been calculated not just for initial conditions stable for up to \num{1e6} turns, but also for cases with shorter stability times. Specifically, a minimum stability of \num{1e2} turns for $\mathrm{FLI}^{\mathrm{WB}}$ and \num{1e4} turns for REM has been assumed.


\section{Analysis of the beam dynamics using the Lyapunov time}\label{sec:times}
The Lyapunov time $T_\mathrm{L}$, defined as the reciprocal of the maximum Lyapunov exponent (MLE)~\cite{Tancredi_2001}, serves as an indicator of the timescale for detecting chaotic dynamics. The MLE can be estimated directly using the dynamic indicators $\mathrm{FLI}/n$ or $\mathrm{FLI}^{\mathrm{WB}}$, both of which offer a straightforward calculation of the MLE. Alternatively, the MLE can be derived via REM, as it corresponds to the exponential growth of REM over time in scenarios with chaotic initial conditions. 

Within the field of accelerator lattice design and the study of non-linear beam dynamics, the DA is commonly used as a figure of merit, calculated using numerical simulations performed using the maximum number of turns $n_\mathrm{max}$.  DA is associated with the stability time $T_\mathrm{S}$, defined as the number of turns required for an orbit to achieve a specified control amplitude. Using fundamental pillars of the field of Hamiltonian dynamical systems, such as KAM theory~\cite{KAM1,KAM2,KAM3} and the Nekhoroshev theorem~\cite{Nekhoroshev:1971aa,Nekhoroshev:1977aa, Bazzani:1990aa,Turchetti:1990aa}, scaling laws with extrapolation capabilities were proposed to describe the long-term evolution of the DA~\cite{invlog,da_and_losses,Bazzani:2019csk}, the most recent model being given by
\begin{equation}
    D(n)=\rho_\ast \times \frac{1}{\left[-2 \mathrm{e} \lambda \mathcal{W}_{-1}\left(-\frac{1}{2 \mathrm{e} \lambda}\left(\frac{\rho_*}{6}\right)^{1 / \kappa}\left(\frac{8}{7} n\right)^{-1 /(\lambda \kappa)}\right)\right]^\kappa} \, ,
    \label{eq:da-scale-law}
\end{equation}
where $\mathcal{W}$ is the Lambert $\mathcal{W}$ function~\cite{Corless1996,10.5555/1403886}. The variables $\rho_\ast$, $\kappa$, and potentially $\lambda$, unless set to $1/2$ following Nekhoroshev original analytical approximation, are free parameters. Here, $\rho_\ast$ is the apparent radius of convergence for the asymptotic series that converts the non-linear Hamiltonian system into an integrable Normal Form, predominantly linked to the strength of the system's non-linearity intensity. $\kappa$ is linked to the number of degrees of freedom, or the phase-space dimension. The free parameters are determined by fitting the DA model to the results of numerical simulations. 

The endeavour to link $T_\mathrm{L}$ with $T_\mathrm{S}$ continues to pose a challenge investigated in diverse branches of physics, including mathematical physics and astrophysics (see, e.g.~\cite{Morbidelli1995} and related works). $T_\mathrm{L}$ is calculated from the MLE and signifies the duration over which an orbit loses the memory of its initial state, with decorrelation primarily in the angle variables. In contrast, $T_\mathrm{S}$ results from a computation reminiscent of a first-passage problem, i.e. the direct evaluation of when the orbit of a given initial condition reaches the absorbing boundary condition and is related to the evolution of the action variables. Consequently, $T_\mathrm{S}$ holds information about a diffusion coefficient that controls the dynamics of these actions, and this approach has been successfully and effectively advanced in recent research~\cite{bazzani2020diffusion,our_paper9,Montanari:2728138,montanari:ipac22-mopost043,montanari:2025}. It is conjectured that the diffusive process occurs in phase-space regions where MLE is positive, and hence $T_\mathrm{L}$ is finite, and generally $T_\mathrm{L} \ll T_\mathrm{S}$. This illustrates the importance of analysing the SR identified by the Lyapunov and stability times and verifying if the Nekhoroshev-like scaling law applies to both domains. The DA calculation, which uses $T_\mathrm{S}$, is constrained by the bound $T_\mathrm{S} \leq n_\mathrm{max}$, with $n_\mathrm{max}$ being the maximum spins in the simulation. Thus, for initial conditions where $T_\mathrm{S} = n_\mathrm{max}$, the true stability duration remains uncertain, only allowing the conclusion that $T_\mathrm{S} \geq n_\mathrm{max}$, which means that the known information provides just a lower bound for the actual stability time. Establishing a direct connection between the Lyapunov time and the stability time could enable the definition of a stability time for stable particles from tracking simulations using the Lyapunov time. It should be emphasised that a general relationship cannot typically be established, as evidenced by numerous cases in the literature, such as stable chaos~\cite{Milani1992, MILANI199713} or intermittency~\cite{sym15061195}. However, even a statistical relationship would be extremely beneficial.

\subsection{Comparison of the properties of the SR based on $T_\mathrm{L}$ and of the DA}
\begin{figure}
    \centering
    \includegraphics[width=\textwidth]{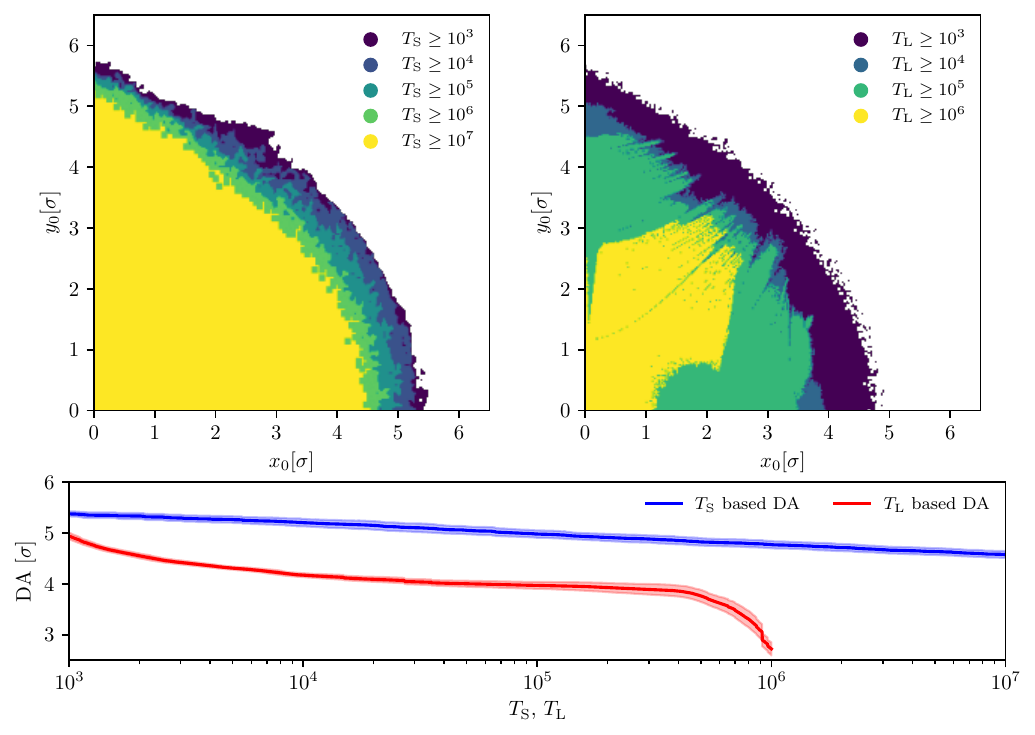}
    \caption{Top: DA, determined using $T_\mathrm{S}$, (left) and SR, determined using $T_\mathrm{L}$, (right). The boundary of the largest connected component of the stability domain is shown as a function of time. The area of the largest connected component is used to compute the equivalent circular radius, which represent the corresponding DA for a given value of time. The very different geometries of the domains related with $T_\mathrm{S}$ or $T_\mathrm{L}$ are clearly visible. Bottom: evolution of the DA based on $T_\mathrm{S}$ or $T_\mathrm{L}$, a sudden drop in dynamic aperture near \(T_\mathrm{L}=\num{1e6}\) is observed. This feature is due to the irregular erosion of the SR clearly seen in the upper-right plot.}
    \label{fig:da_example}
\end{figure}

The determination of DA is performed using data from the $300\times300$ Cartesian ROI depicted in Fig.~\ref{fig:seed_presentation}. Analysing this grid of initial conditions, the variation of the largest connected component of the SR is assessed at varying $n$ values, up to a value of $10^7$ turns. The area of each SR is calculated and used to derive the radius of an equivalent area circle, representing $D(n)$ as per Eq.~\eqref{eq:da-scale-law}. The error in this method arises from the grid step size defined for the initial conditions. An analogous method is applied to evaluate similar metrics using $T_\mathrm{L}$. Figure~\ref{fig:da_example} visually depicts this process, showing the SR related to the stability time (left) and the Lyapunov time (right), along with the evolution of the DA and the counterpart determined from $T_\mathrm{L}$. Noticeable differences in the SR geometries corresponding to the two times are evident. In particular, the geometry becomes highly irregular for the SR tied to Lyapunov time with increasing $T_\mathrm{L}$. This irregularity is reflected in the equivalent radius evolution in the lower plot. Unlike DA, which exhibits a fairly smooth progression, a pronounced decrease is evident as \num{1e6} turns are approached for the case based on Lyapunov time.

\begin{figure}
    \centering
    \includegraphics[width=\textwidth]{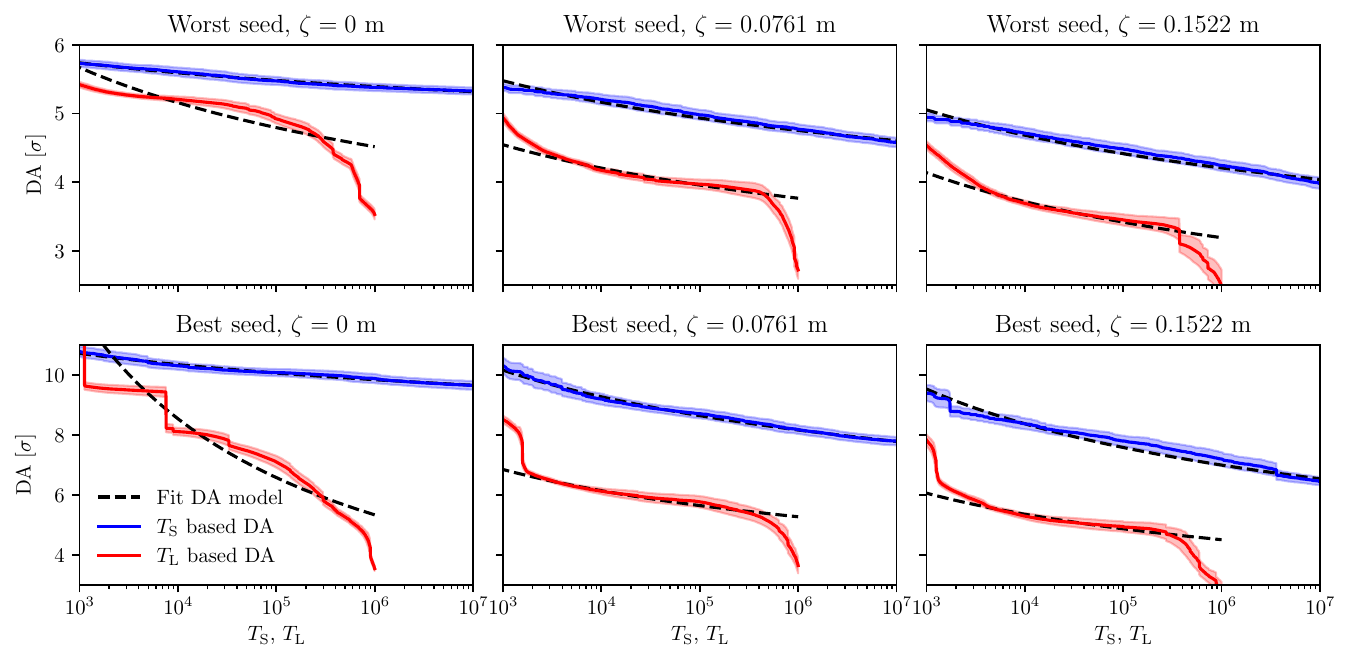}
    \caption{Comparison between the DA evaluated on $T_\mathrm{S}$ (blue lines) and $T_\mathrm{L}$ (red lines) evaluated using $\mathrm{FLI}^\mathrm{WB}$ for the two seeds and the three values of $\zeta_0$ considered. A fit of the scaling law~\eqref{eq:da-scale-law} is also shown (dotted lines), whose free parameters are reported in Table~\ref{tab:da_comparison}. The agreement between model and numerical data is better for the DA based on $T_\mathrm{S}$, although when $T_\mathrm{L}$ is used the agreement with the model improves as a function of $\zeta_0$.}
    \label{fig:da_comparison}
\end{figure}

Figure~\ref{fig:da_comparison} provides a detailed comparison of the DA calculated using $T_\mathrm{S}$ and $T_\mathrm{L}$ for two seeds and three values of $\zeta_0$. Overall, the standard DA yields smoother curves compared to those based on Lyapunov time. This indicates that the average over the angle in the $x_0-y_0$ space is more efficient when the analysis is based on the stability time. In particular, when $\zeta_0 \neq 0$, the two curves appear nearly parallel, whereas for $\zeta_0 = 0$, the curves intersect at lower values of the time variable. This pattern becomes even more pronounced when the model is fitted to the numerical data. The construction of the model involves fitting the parameters $\kappa$ and $\rho_\ast$ to the observed data. As mentioned previously, $\kappa$ depends on the dimension of the phase space and is thus expected to be the same for all cases in which longitudinal dynamics is fully included, while a different value is expected when $\zeta_0 = 0$.

The fit parameters obtained are shown in Table~\ref{tab:da_comparison}, together with the calculated reduced chi-squared $\chi^2_\nu$. The uncertainty in the fit parameters is estimated by observing the variation when the fit is applied to a DA curve adjusted within its uncertainty band, compared to the uncertainty in the fit parameters taken as half of their variation range. For the DA case, the fit is of better quality when $\zeta_0=0$, regardless of the seed, and $\rho_\ast$ appears to be independent of $\zeta_0$ when this variable is different from zero. Furthermore, the fit parameters increase by approximately a factor of two between the case of zero longitudinal variable and non-zero values. 

\begin{table}
    \centering
    \caption{Fitted parameters for the scaling law~\eqref{eq:da-scale-law} applied to the SR determined from $T_\mathrm{S}$ and $T_\mathrm{L}$ for the cases of the two seeds and three values of $\zeta_0$. The free parameters $\kappa$ and $\rho_\ast$ are fitted independently for each individual case with $\zeta_0=0$. When $\zeta_0 \neq 0$ the results for the two values of $\zeta_0$ are fitted together to impose the same value of $\kappa$. The value of $\lambda$ is fixed to $1/2$ for all fits. The reduced chi-squared $\chi^2_\nu$ is also reported.}
    \begin{tabular}{cl|ccc|ccc}
        \toprule
        & & \multicolumn{3}{c|}{Fit of $T_\mathrm{S}$ based DA} & \multicolumn{3}{c}{Fit of $T_\mathrm{L}$ based DA} \\
        Seed & $\zeta_0$ & $\rho_\ast$ & $\kappa$ & $\chi^2_\nu$ & $\rho_\ast$ & $\kappa$ & $\chi^2_\nu$ \\
& [\si{\meter}] & & & & & & \\
        \midrule
        \multirow{3}{*}{Best}
        & $0.0$& $21.4 \pm 0.2$ & $0.121 \pm 0.001$ & 0.03 &
                $579 \pm 3$ & $1.108 \pm 0.009$ & 4.7 \\
         & $0.0761$ & $45.2 \pm 1.9$ & \multirow{2}{*}{$0.31 \pm 0.01$} & \multirow{2}{*}{0.26} &
                 $63 \pm 9$ & \multirow{2}{*}{$0.50 \pm 0.05$} & \multirow{2}{*}{0.29} \\
         & $0.1522$ & $40.3 \pm 1.4$ &  &  &
                 $55 \pm 8$ &  &  \\
        \midrule
        \multirow{3}{*}{Worst}
        & $0.0$& $9.9 \pm 0.2$ & $0.091 \pm 0.001$ & 0.03 &
                $27 \pm 1$ & $0.33 \pm 0.01$ & 5.01 \\
         & $0.0761$ & $16.9 \pm 0.4$ & \multirow{2}{*}{$0.217 \pm 0.007$} & \multirow{2}{*}{0.26} &
                 $24 \pm 3$ & \multirow{2}{*}{$0.35 \pm 0.04$} & \multirow{2}{*}{0.21} \\
         & $0.1522$ & $15.1 \pm 0.3$ &  &  & 
                 $22 \pm 3$ &  & \\
        \bottomrule
    \end{tabular}
    \label{tab:da_comparison}
\end{table}

In the case of the $T_\mathrm{L}$ based DA, the quality of the fit is much worse for $\zeta_0=0$, while in the other two cases the quality is comparable with that of the fit for the standard DA. These results suggest that when non-linear contributions are sufficiently strong, the DA computed using $T_{\mathrm{S}}$ and $T_{\mathrm{L}}$ exhibits good agreement. This indicates that DA based on Lyapunov time might serve as a viable alternative tool for analysis in such cases. However, when the system lacks the necessary presence of chaotic structures, this relationship breaks down. In these conditions, the agreement between $T_{\mathrm{S}}$ and $T_{\mathrm{L}}$ deteriorates, as well as the underlying assumptions required to observe the strong evolution of DA over time.

\subsection{Direct relationship of $T_\mathrm{L}$ and $T_\mathrm{S}$}
We have also studied the possible direct correlation between $T_\mathrm{L}$ and $T_\mathrm{S}$ as a function of the initial radial amplitude. The Lyapunov time is an estimate of the exponential loss of correlation between the evolution of two orbits with arbitrarily close initial conditions. This means an exponential increase of the distance of the orbits that is usually due to the phase variables and not action variables, if one considers a weakly-chaotic region resulting from the overlapping of non-linear resonances. The stability time is related to the increase in the amplitude of the orbits up to a fast instability threshold. If the dynamics of actions can be described by a diffusion equation coming from a perturbation, then $T_\mathrm{S}$ is inversely proportional to the diffusion coefficient, which is estimated by the square of the strength of the perturbation, whereas if we assume a ballistic increase of the action variables, the stability time is inversely proportional to the perturbation. For weakly-chaotic systems, the stability time $T_\mathrm{S}$ is mainly related to the action diffusion in the phase space, whereas the Lyapunov exponent is generated by the decorrelation on the angle variables. The relation between $T_\mathrm{L}$ and $T_\mathrm{S}$ should consider how angle decorrelation can induce diffusive behaviour in action variables and the dependence of stability time on the extension of the chaotic region. This relation is discussed in~\cite{Morbidelli1995}, where the authors consider different regimes to justify a power-law relation $T_\mathrm{S} \sim \alpha T_\mathrm{L}^\beta$ or an exponential relation $T_\mathrm{S} \sim \alpha \exp(\beta T_\mathrm{L})$, where $\alpha, \beta$ are the free parameters. Their results suggest that there is no universal relationship between $T_\mathrm{S}$ and $T_\mathrm{L}$, since the actual regime depends on the specific characteristics of the Hamiltonian system under study.

Extensive numerical simulations have been performed to study the correlation between $T_\mathrm{S}$ and $T_\mathrm{L}$, as functions of the initial radius $r_0 = \sqrt{x_0^2 + y_0^2}$.
We evaluated the averages $T_\mathrm{S}$ and $T_\mathrm{L}$ for a set of initial conditions in the amplitude interval $[r_0-\Delta r/2, r_0 +\Delta r/2]$ and in the angular sector comprised between $[0,\pi/2]$. The averages of $T_\mathrm{S}$ and $T_\mathrm{L}$ are calculated for each interval and the resulting values are plotted as a function of $r_0$. We consider $T_\mathrm{S}$ computed using numerical simulations performed up to \num{1e7} turns, while $T_\mathrm{L}$ is evaluated by $\mathrm{FLI}^{\mathrm{WB}}$ at \num{1e6} turns.

\begin{figure}
    \centering
    \includegraphics[width=1\textwidth]{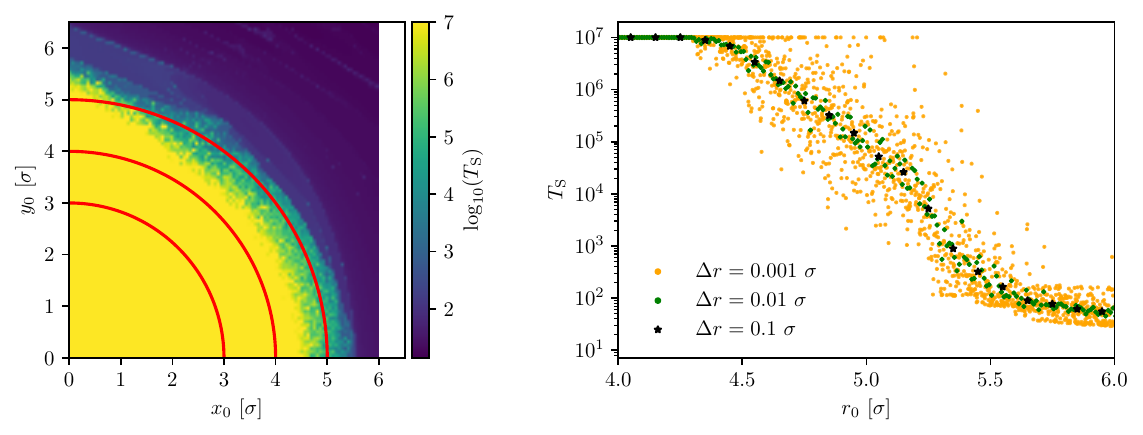}
    \caption{Left: stability domain based on $T_\mathrm{S}$ together with three circles used to evaluate the average value in different amplitude intervals. Right: mean values of $T_\mathrm{S}$ as a function of $r_0$ for three values of $\Delta r$. The best compromise between statistical fluctuations, due to $\Delta r$ being too small, and loss of information, due to $\Delta r$ being too large, is achieved for $\Delta r = 0.01\sigma$. These results refer to the case of the worst seed and $\zeta_0 = \SI{0.0761}{m}$. The same approach is applied to the analysis of $T_\mathrm{L}$.}
    \label{fig:ts_vs_r0}
\end{figure}

To find the optimal choice $\Delta r$, we first calculate the average values of $T_\mathrm{S}$ and $T_\mathrm{L}$ for various options for $\Delta r$ and compare their resulting curves directly. The optimal $\Delta r$ is the one that strikes the best balance between statistical fluctuations and the preservation of information. Figure~\ref{fig:ts_vs_r0} illustrates the outcome of the analysis for $T_\mathrm{S}$, with $T_\mathrm{L}$ producing similar results, pertaining to the worst seed of the HL-LHC lattice. It is evident that the optimal balance occurs at $\Delta r = 0.01\sigma$, which is the value selected for subsequent analysis. For the uncertainty in the average values of $T_\mathrm{S}$ and $T_\mathrm{L}$, we account for the standard deviation of their averages over each interval.

\begin{figure}
    \centering
    \includegraphics[width=1\textwidth]{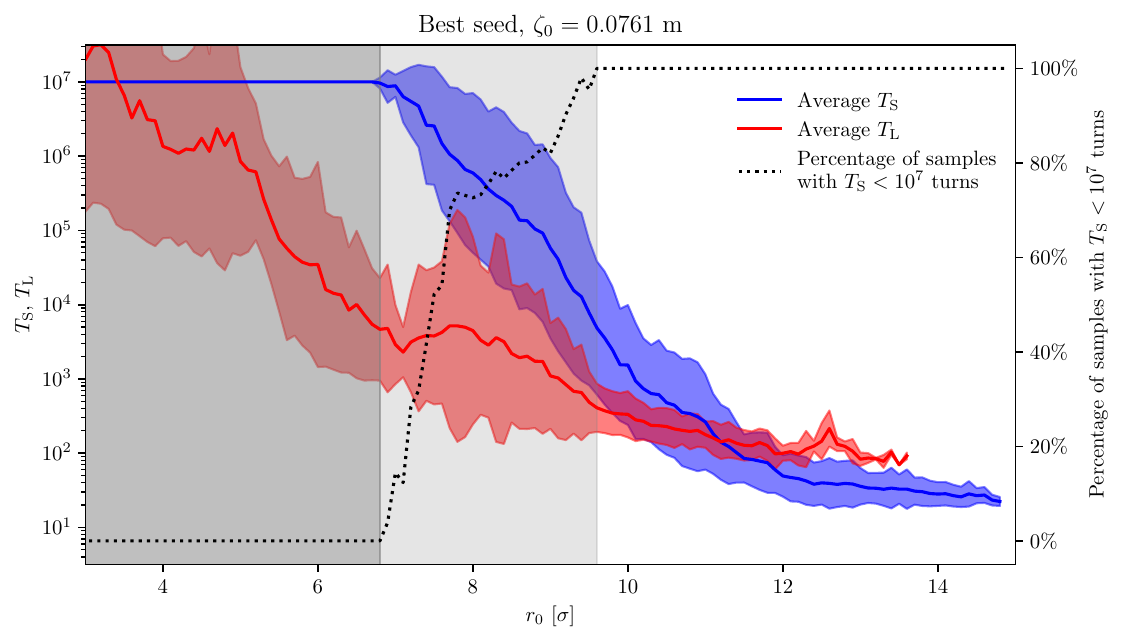}
    \caption{Average values of $T_\mathrm{L}$ and $T_\mathrm{S}$ as a function of $r_0$ for $\Delta r = 0.01\sigma$, for the best seed and $\zeta_0 = \SI{0.0761}{\meter}$. $T_\mathrm{L}$ is calculated using $\mathrm{FLI}^\mathrm{WB}$ at \num{1e6} turns. The uncertainties on the mean values are represented by the standard deviation of $T_\mathrm{S}$ and $T_\mathrm{L}$ in each interval. The black dotted line indicates the fraction of samples for which $T_\mathrm{S} < \num{1e7}$ turns as a function of $r_0$. The region shaded in dark grey represents the case for which all samples feature $T_\mathrm{S}=\num{1e7}$ turns; the region shaded in lighter grey represents the case in which some samples feature $T_\mathrm{S} < \num{1e7}$ turns; the region shaded in white represents the case in which all samples feature $T_\mathrm{S} < \num{1e7}$ turns, and this is the region where the fit is performed (see Fig.~\ref{fig:morbidelli}).}
    \label{fig:ts_vs_tl}
\end{figure}

When comparing directly $T_\mathrm{L}$ and $T_\mathrm{S}$, as shown in Fig.~\ref{fig:ts_vs_tl}, we can observe that the two curves exhibit a qualitative similar behaviour with respect to the initial amplitude $r_0$ with a fast increase in stability time up to a saturation value when $r_0$ is below a certain value. Clearly, the saturation effect is missing for the curve that refers to the Lyapunov time. 

We note that average values close to the saturation value of $T_\mathrm{S}$ corresponding to \num{1e7} turns are going to be affected by a bias due to the initial conditions that are stable up to the maximum number $n_\mathrm{max}$. 
The extent of this bias can be observed in Fig.~\ref{fig:ts_vs_tl}, which shows the percentage of samples at different values of $r_0$ that have not reached $n_\mathrm{max}$. To cope with this problem, only the subset of data corresponding to the white region, i.e. corresponding to the case with $T_\mathrm{S} \leq \num{1e7}$ turns, has been considered. 

\begin{figure}
    \centering
    \includegraphics[width=1\textwidth]{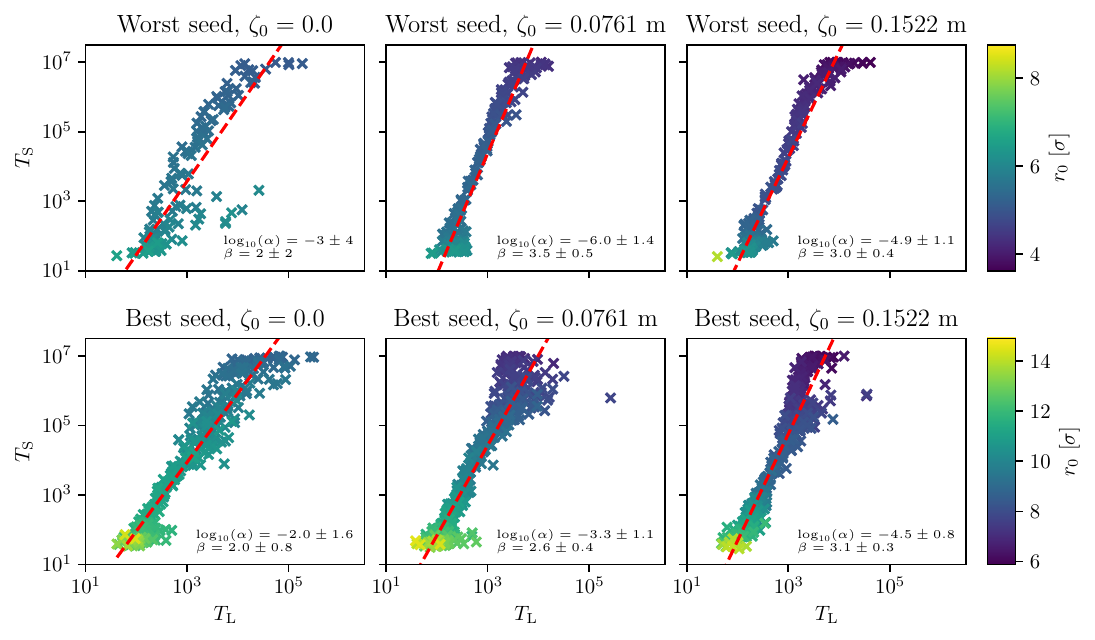}
    \caption{Correlation plots between $T_\mathrm{S}$ and $T_\mathrm{L}$ for the two seeds and the three values of $\zeta_0$ considered. $T_\mathrm{L}$ is evaluated using $\mathrm{FLI}$ at $n=\num{1e6}$ turns. The red line represents the best fit of the data in region shaded in white in Fig.~\ref{fig:ts_vs_tl}, using a power-law model $T_\mathrm{S} = \alpha  T_\mathrm{L}^\beta$ typical of the resonance-overlapping regime~\cite{Morbidelli1995}.}
    \label{fig:morbidelli}
\end{figure}

\begin{figure}
    \centering
    \includegraphics[width=0.7\linewidth]{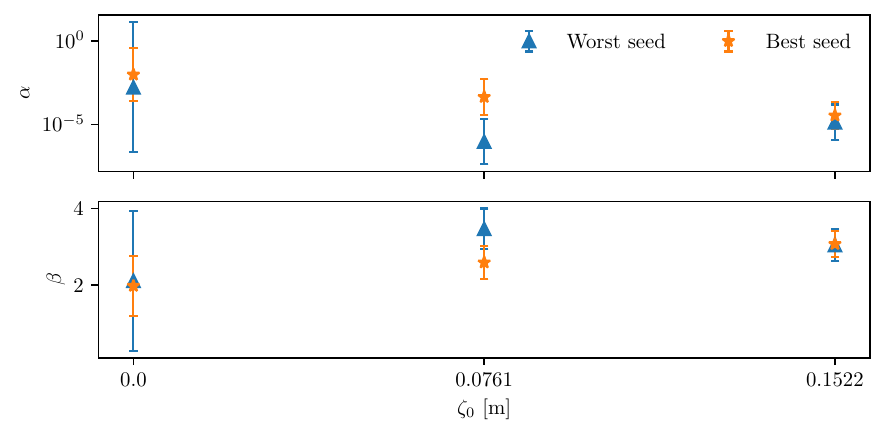}
    \caption{Overview of the fit parameters of the power law between $T_\mathrm{S}= \alpha T_\mathrm{L}^\beta$ presented in Fig.~\ref{fig:morbidelli}. The parameters for the worst and best seeds are compatible while the dependence on $\zeta_0$ is clearly visible.}
    \label{fig:morbielli-fit parameters}
\end{figure}

In Fig.~\ref{fig:morbidelli}, we show the correlation between the averages, computed at various $r_0$ values, of $T_\mathrm{S}$ and $T_\mathrm{L}$ for the two seeds and three values of $\zeta_0$ for the HL-LHC lattice. We observe that the stability time exhibits a clear correlation in a log-log scale plot, regardless of the value of $r_0$. To inspect the power-law correlation, which corresponds to the resonance-overlapping regime~\cite{Morbidelli1995}, we fit the data using a power law of the form $T_\mathrm{S} = \alpha T_\mathrm{L}^\beta$, where $\alpha$ and $\beta$ are the free parameters. The resulting fit is also presented in Fig.~\ref{fig:morbidelli} (red dashed lines), together with the values of the fit parameters. We note that, although the power-law fit is computed over a limited range of amplitudes, it agrees well with the complete set of data reported in the plots. The values of the parameters $\alpha$ and $\beta$ are shown in Fig.~\ref{fig:morbielli-fit parameters} as a function of $\zeta_0$. The parameters for the two seeds are fully compatible, i.e. they are, to a large extent, independent of the seed. A very slight dependence on $\zeta_0$ might be observed.

The power law is capable of capturing the correlation between $T_\mathrm{S}$ and $T_\mathrm{L}$ with the exponent $\beta\in [2,4]$, suggesting that the action dynamics could have features compatible with diffusion regimes.

This encouraging outcome indicates that the power law could serve as an approximation of the data, potentially providing extrapolation abilities that merit exploration in accelerator lattice optimisation. Nevertheless, it is important to recognise that these findings can be greatly influenced by the system, suggesting that the power law's extrapolation abilities need further examination across various scenarios and should be approached carefully.

\section{Conclusions} \label{sec:conc}
Chaos indicators serve as an effective means for analysing the phase-space characteristics of a dynamical system, offering insights into chaotic regions and their boundaries while maintaining a low computational cost. This study investigates the application of advanced chaos indicators on realistic HL-LHC lattices to evaluate their usage in accelerator physics and their applicability in optimising accelerator lattices. Specifically, we concentrated on the Fast Lyapunov Indicator with Birkhoff weights ($\mathrm{FLI}^{\mathrm{WB}}$) and the Reverse Error Method (REM), considering their promising results as demonstrated in prior research on symplectic maps.

This study compares the performance of the proposed indicators for identifying chaotic behaviour, which is indicated by a positive MLE. This method effectively highlights structures from non-linear resonances and outlines the boundary of the chaotic layer. It offers a broader application than the FMA method and can be used for realistic accelerator models, like those of the HL-LHC, incorporating synchrotron dynamics. Detecting chaos enables us to pinpoint phase-space regions where action dynamics may become unstable after a certain number of turns, thereby supporting the use of the functional relationship between stability time and the Nekhoroshev optimal estimate for the perturbative series.

We propose that the presence of a positive MLE in a significant portion of phase space validates a diffusion model for the dynamics, enabling the possibility of averaging over fast variables (specifically, the angles) to understand how orbits drift to the boundary of the SR, at least on average. To explore the link between chaos indicators and orbit stability, we analysed the average stability duration alongside the average Lyapunov time at varying amplitudes, employing an angular averaging process akin to that for determining the DA scaling laws. The analysis revealed a distinct power-law relationship between these two times, reminiscent of the resonance-overlap regime outlined in~\cite{Morbidelli1995}. This indicates that the action dynamics may exhibit a combination of ballistic and diffusion behaviours. Moreover, the power law may offer a reliable extrapolation to higher $T_\mathrm{S}$ values, which are critical to optimising accelerator lattices.

Inspired by advanced software frameworks that allow efficient GPU parallel tracking of multiple initial conditions, we proceeded with a statistical analysis of the correlation between the conventional SR defined by stability time, i.e. the DA, and the region defined by Lyapunov time. We discovered that both measures follow a similar scaling law, effectively described by a Nekhoroshev-like scaling law. This suggests that both approaches to defining a SR may exhibit analogous long-term behaviour. Consequently, the SR based on Lyapunov time might serve as a substitute for the SR based on stability time or the DA. This finding is encouraging because it implies that Lyapunov time can be effectively used to examine the dynamic aperture of an accelerator lattice without executing a comprehensive stability analysis, thereby reducing computational expense.

In conclusion, the encouraging outcomes presented in this study may stimulate a more in-depth investigation into the applicability of chaos indicators within accelerator physics, specifically aimed at optimising accelerator lattices. The objective would be to broaden and implement these methods on realistic accelerator lattices, which include long-range beam-beam interactions. Considering the vast parameter space that requires substantial computational power, the enhanced performance offered by the chaos indicator may prove crucial for conducting thorough examinations of lattice characteristics while cutting down computational expenses.
\section*{Data availability}
Data sets generated during the current study are available from the corresponding author on reasonable request. %
\appendix
\section{Overview of chaos indicators} \label{app:overview}
We recall the fundamental theory and concepts of our selection of dynamic indicators. For a complete description of the individual indicators, we encourage the reader to refer to the sources cited, e.g.~\cite{skokos2016chaos, Skokos2010b}, and references therein. A similar selection of dynamic indicators was also explored in~\cite{PhysRevE.107.064209}, with an application to a modulated Hénon map with octupolar kicks, for which a full analytical expression of the tangent map is available and tracking up to physically relevant time scales is possible.

Let us consider a non-autonomous Hamiltonian map $M(\vb{x}, n)$, defined in $\mathbb{R}^{2 d}$, where $d$ is the number of degrees of freedom and $n$ is the discrete time. We denote by $DM(\vb{x}, n)$ the symplectic Jacobian matrix $(DM)_{ij} = \partial M_i / \partial x_j$ evaluated at the point $\vb{x}$ at time $n$.

The orbit of the map $\vb{x}_n$ for an initial condition $\vb{x}$, and the recurrence for the tangent map $\mathrm{L}_{n}$ are given by
\begin{equation}
    \begin{aligned}
    &\mathbf{x}_{n}=M\left(\mathbf{x}_{n-1}, n-1\right) \qquad& \mathbf{x}_{0}&=\mathbf{x} \\
    &\mathrm{L}_{n}(\mathbf{x})=D M\left(\mathbf{x}_{n-1}, n-1\right) \mathrm{L}_{n-1}(\mathbf{x}) \qquad& \mathrm{L}_{0}&=\mathrm{I} \, .
    \end{aligned}
\end{equation}

The chaotic character of an initial condition is thus probed by using the linear response of the system to different combinations of stochastic deviations via dynamic indicators.

\subsection{Fast Lyapunov Indicator}
The Fast Lyapunov Indicator (FLI)~\cite{Froeschle1997} is one of the most established Lyapunov-based dynamic indicators, due to its straightforward implementation and sensitivity to identify chaotic structures, such as the Arnold web~\cite{Lega2016fli}. It accomplishes this by providing a rapid numerical estimate of the MLE.

The FLI dynamic indicator considers the linear response of the system when subjected to small stochastic perturbations of the initial condition $\vb{x}$, specifically $\vb{y}_0=\vb{x}+\epsilon\boldsymbol{\xi}$, where $\boldsymbol{\xi}$ is a random vector with a zero mean and a unit covariance matrix. The linear response vector, denoted as $\boldsymbol{\Xi}_{n}(\mathbf{x})$, is defined as:

\begin{equation}
    \boldsymbol{\Xi}_{n}(\mathbf{x})=\lim _{\epsilon \rightarrow 0} \frac{\mathbf{y}_{n}-\mathbf{x}_{n}}{\epsilon}= D M\left(\mathbf{x}_{n-1}, n-1\right) \boldsymbol{\Xi}_{n-1} = \mathrm{L}_{n}(\mathbf{x}) \boldsymbol{\xi} \,.
    \label{eq:linear_response}
\end{equation}

The FLI at the nth iteration for a perturbed initial condition $\vb{x}$ is given by:

\begin{equation}
    \text{FLI}_n(\vb{x}) = \log\norm{\boldsymbol{\Xi}_{n}(\vb{x})}\,.
    \label{eq:base_fli}
\end{equation}

The ratio $\mathrm{FLI}_n(\vb{x})/n$ converges to the MLE as $n$ tends to infinity. For initial conditions associated with regular orbits, where the largest Lyapunov exponent is zero, $\mathrm{FLI}_n(\vb{x})/n$ follows a power-law convergence to zero. 

The numerical evaluation of FLI depends on the initial choice of $\boldsymbol{\xi}$, which can lead to the detection of different spurious structures in phase space~\cite{PhysRevE.107.064209}. To address this problem, one considers the statistical properties of multiple $\boldsymbol{\xi}$ displacements or employs more advanced indicators such as the orthogonal fast Lyapunov Indicator (OFLI)~\cite{Barrio2016}, which allows one to examine the full spectrum of Lyapunov exponents. Alternatively, one can opt for an invariant definition of the dynamic indicator, such as the Lyapunov Error (LE)~\cite{Panichi2018}, which provides a Lyapunov-related dynamic indicator independent of the choice of the initial perturbation.

If an analytic expression for $DM(\vb{x}, n)$ is not available, estimating the linear response can be performed by the shadow-particle method~\cite{Skokos2010b}. This method is based on the tracking of two nearby initial conditions $\vb{x}_n$ and $\vb{y}_n$, evaluating the estimate $\boldsymbol{\Xi}_{n}(\mathbf{x})=(\mathbf{y}_{n}-\mathbf{x}_{n})/\epsilon$ where $\epsilon$ is the initial distance. This is performed by scaling
the displacement vector $\vb{y}_n - \vb{x}_n$ every $\tau$ iterations to the initial length $\epsilon$ and computing the product of the scaling factors. Such a procedure prevents spurious effects in the FLI evaluation arising from excessive orbit drifting. For clarity, this periodically modified $\vb{y}$ is denoted as $\vb{y}'$, and its formulation is as follows:

\begin{equation}
\begin{aligned}
    \vb{y}'_0 &= \vb{y}_0 \,; \\
    \vb{y}'_i &= DM(\vb{x}_{i\tau - 1}, i\tau - 1)\cdots DM(\vb{x}_{(i-1)\tau}, (i-1)\tau)\left[\vb{x}_{(i-1)\tau}+\epsilon\,\frac{\vb{y}'_{i-1}-\vb{x}_{(i-1)\tau}}{\norm{\vb{y}'_{i-1}-\vb{x}_{(i-1)\tau}}}\right]\,, \qquad 0 \leq i \leq n/\tau \,.
\end{aligned}
\end{equation}

To incorporate $\vb{y}_n'$ into the FLI expression, we can take advantage of the properties of the logarithm in Eq.~\eqref{eq:base_fli}. This allows us to express $\mathrm{FLI}_n(\vb{x})/n$ as an average along the trajectory $\vb{x}_n$, resulting in:

\begin{equation}
    \frac{\text{FLI}_n(\vb{x})}{n} = \frac{1}{n}\sum_{i=1}^{n/\tau} \log{\norm{\frac{\vb{y}_{i}' - \vb{x}_{i\tau}}{\epsilon}}}\,.
    \label{eq:fli_mean}
\end{equation}

It should be noted that it has been emphasised~\cite{Tancredi_2001} how the choice of both the value of $\epsilon$ and the time interval $\tau$ between displacement resets to $\epsilon$ can significantly influence the final evaluation of $\boldsymbol{\Xi}_{n}(\mathbf{x})$ and, consequently, $\mathrm{FLI}_n(\vb{x})$. This dependence on the choice of $\epsilon$ and $\tau$ implies the need to perform a convergence study for each specific application of FLI, to ensure that the chosen values of $\epsilon$ and $\tau$ are suitable for the problem at hand.
\subsection{Birkhoff Weighted Fast Lyapunov Indicator}
In the case of the quasi-periodic time series, such that associated to regular orbits in the phase space, the use of non-uniform weights to compute an average quantity can yield superior convergence properties compared to a standard uniform average~\cite{Das_2018}. In the work of Das et al.~\cite{Das_2017}, it is presented how employing the Weighted Birkhoff averaging method $\mathrm{WB}_n$ to evaluate the maximum Lyapunov exponent provides accelerated convergence rates. Furthermore, improvements in FLI evaluation convergence rates were observed in cases involving a modulated Hénon map with octupolar kicks~\cite{PhysRevE.107.064209}.

The Weighted Birkhoff averaging applies a weighting function $w\left(\frac{i}{n}\right)$ to the underlying data, akin to a window function in frequency space. An especially effective weighting function in enhancing the convergence of quasi-periodic time series averages~\cite{Das_2018} is:

\begin{equation}
    w(t):= 
    \begin{cases}
        \exp \left[-\frac{1}{t(1-t)}\right], & \text { for } t \in(0,1) \\ 
        0, & \text { for } t \notin(0,1)
    \end{cases} \,.
    \label{eq:birkhoff}
\end{equation}

Substituting the standard mean with $w(t)$ in Eq.~\eqref{eq:fli_mean} we get:

\begin{equation}
    \text{FLI}_n^{\mathrm{WB}}(\vb{x}) = \frac{1}{\tau}\sum_{i=0}^{n/\tau} w\left(\frac{i}{n/\tau}\right) \log{\norm{\frac{\vb{y}_{i}' - \vb{x}_{i\tau}}{\epsilon}}} \ ,
    \label{eq:fli_birkhoff}
\end{equation}
and in scenarios where the series $\log{\norm{\frac{\vb{y}_{i}' - \vb{x}_{i\tau}}{\epsilon}}}$ exhibits quasi-periodic behaviour, that is, non-chaotic behaviour, $\mathrm{FLI}_n^{\mathrm{WB}}(\vb{x})$ will converge more rapidly than $\mathrm{FLI}_n(\vb{x})/n$.

\subsection{Reverse Error Method}
The Reverse Error Method (REM)~\cite{Panichi2016, Panichi2017, Panichi2018, Panichi19} is based on the linear response of a dynamical system (susceptibility) to small stochastic perturbations applied over $n$ iterations of the map $M$, followed by $n$ iterations of the inverse map $M^{-1}$. This process can be described as follows:
\begin{equation}
\begin{aligned}
\mathbf{y}_{n^{\prime}}&=M\left(\mathbf{y}_{n^{\prime}-1}, n^{\prime}-1\right)+\epsilon \boldsymbol{\xi}_{n^{\prime}} \quad &1 &\leq n^{\prime} \leq n \\
\mathbf{y}_{n^{\prime}}&=M^{-1}\left(\mathbf{y}_{n^{\prime}-1}, 2 n-n^{\prime}\right)+\epsilon \boldsymbol{\xi}_{n^{\prime}} \quad &n+1 &\leq n^{\prime} \leq 2 n \, ,
\end{aligned}
\end{equation}
where $\vb{y}_0 = \vb{x}$. The linear response at iteration $2n$ is denoted as:
\begin{equation}
\boldsymbol{\Xi}_{\mathrm{R} n}(\mathbf{x})=\lim _{\epsilon \rightarrow 0} \left \langle \frac{\mathbf{y}_{2 n}-\mathbf{x}}{\epsilon}\right \rangle \, ,
\end{equation}
and the REM dynamic indicator is defined as:
\begin{equation}
\Bigl(\text{REM}_n(\vb{x})\Bigr)^2 = \frac{1}{ 2} \,\, \left (\frac{\vb{y}_{2n}-\vb{x}}{ \epsilon}\right )^2\, ,
\label{eq2_21}
\end{equation}
which represents the Euclidean distance between the initial condition and the resulting displaced particle after the forward and backward tracking in the limit of zero noise amplitude averaged on the noise realisations.

The standard approach in REM treats numerical round-off as a form of pseudo-random deviation along the orbit, with an amplitude typically on the order of $\epsilon \sim 10^{-16}$ in the 8-byte standard IEEE754~\cite{8766229} representation of real numbers. This round-off is akin to white noise in terms of its spectral characteristics.

The key aspects of REM include its ease of implementation, particularly when an expression for the inverse map is available, and the result is independent of the choice of the initial perturbation. However, since we consider a single realisation of the round-off noise, we may have relevant fluctuations as a function of the iteration number $n$. 

By treating the numerical round-off error as a white-noise process, it is expected to observe a power-law evolution of REM over time for regular orbits, while for chaotic orbits it is expected to observe an exponential growth determined by their MLE. This difference in growth rates can be exploited to achieve a fast binary classification of regular and chaotic orbits.
\subsection{Frequency Map Analysis}
The Frequency Map Analysis (FMA) examines how the main frequencies of an orbit evolve over time, leveraging the quasi-periodic nature of regular orbits of Hamiltonian systems (i.e., KAM tori). J.~Laskar first introduced the FMA in the field of celestial mechanics, where it quickly spread to other disciplines, yielding significant outcomes, particularly in accelerator physics (see, e.g. the following selected references~\cite{laskar1995frequency,lega1996numerical,papaphilippou1996frequency,papaphilippou1998global,Laskar1999, laskar2000application,PhysRevSTAB.4.124201,1288929,Papaphilippou:PAC03-RPPG007,Laskar2003,PhysRevSTAB.6.114801,shun2009nonlinear,PhysRevSTAB.14.014001,papaphilippou2014,tydecks:ipac18-mopmf057,PhysRevAccelBeams.22.071002}).

Given a Hamiltonian system of $n$ degrees of freedom in action-angle variables $H(I,\theta) = H_0 (I) + \varepsilon H_1 (I, \theta)$, where for $\varepsilon=0$ the Hamiltonian $H_0(I)$ is integrable. 
If the system is non-degenerate
\begin{equation}
    \operatorname{det}\left(\frac{\partial \nu(I)}{\partial I}\right)=\operatorname{det}\left(\frac{\partial^2 H_0(I)}{\partial I^2}\right) \neq 0 \,,
\end{equation}
the application
\begin{equation}
    \begin{array}{r}
    F: \mathbb{R}^{n} \longrightarrow \mathbb{R}^n \\
    (I) \longrightarrow(\nu)
    \end{array}
\end{equation}
is a diffeomorphism between the action and the frequency space. This means that the invariant tori are equally identified by the action variables $I$ or by their corresponding frequency vector $\mathbf{\nu}$. For a non-degenerate system, when $\varepsilon$ is sufficiently small, the KAM theorem~\cite{KAM1, KAM2, KAM3}, asserts that there still exists a set of initial conditions for which the perturbed system still possesses regular motion, that is, KAM tori set, for which, according to Pöschel's theorem~\cite{Poschel1982}, a similar diffeomorphism still exist even if the KAM tori define a Cantor set.

Within this theoretical framework, one distinguishes between regular orbits, which exhibit a discrete Fourier spectrum, namely that of the KAM tori, and chaotic orbits, whose Fourier spectrum is complex since the corresponding orbit does not lie on a torus~\cite{1288929}.

FMA consists in performing numerical evaluations of the frequency vector $\vb{\nu}$ from a time series corresponding to a certain interval $[i, i+2n-1]$, determining the Euclidean distance between the frequencies computed using the data in the time interval $[i,i+n-1]$ and $[i+n, i+2n-1]$, and then inspecting the evolution of the distance as a function of $n$. In case of a regular orbit, the distance tends to zero, the frequency vector converging to the true values to the precision of the numerical method used to determine the orbit frequency. In the case of a chaotic orbit, the distance will stay away from zero, showing a sort of diffusion in the frequency space~\cite{laskar2000application}.

To efficiently estimate numerically the fundamental frequencies of a time series, multiple studies have been performed to improve standard algorithms such as the Fast Fourier Transform (FFT) or the Average Phase Advance (APA)~\cite{laskar1992measure,Laskar1999, Bartolini:316949,bartolini1998computer}. In the work of Bartolini et al.~\cite{Bartolini:292773}, the fundamental frequency is evaluated using an FFT combined with a Hanning filter and an interpolation algorithm. In recent studies~\cite{russo:ipac2021-thpab189}, the APA algorithm is improved by applying the Weighted Birkhoff averaging~\cite{Das_2018}, which will be used in this paper to perform the FMA evaluation. In the context of this research, we will use the APA algorithm improved with the Weighted Birkhoff averaging to estimate the fundamental frequencies.

To compare FMA with the other dynamic indicators, we define FMA$_n$ as the Euclidean distance between two vectors of fundamental frequencies $\nu_1$ and $\nu_2$, evaluated, respectively, over the time intervals $[0, n/2]$ and $[n/2, n]$ for an orbit. An initial condition on a regular torus will exhibit regular motion and have FMA$_n$ converging to zero. In contrast, for an initial condition that is not in a regular torus, the distance between $\nu_1$ and $\nu_2$ is bounded away from zero.

\section{Features of dynamic indicators} \label{app:features}

\subsection{Use of Birkhoff weights with FLI}\label{subsec:lhc_fli_bk}

To quantify the convergence improvements given by the Birkhoff weights, we compare the values obtained for FLI at different times for one of the HL-LHC lattices, using the standard approach that considers the mean in Eq.~\eqref{eq:fli_mean}, that is, $\mathrm{FLI}/n$, or the weighted mean based on the use of Birkhoff weights as in Eq.~\eqref{eq:fli_birkhoff}, that is, $\mathrm{FLI}^{\mathrm{WB}}$.

In this analysis, we consider two ensembles of regular and chaotic particles that have been classified by means of the FLI indicator value computed for $n=\num{1e5}$ turns. Inevitably, this evaluation does not reach an absolute level of reliability due to the limited number of turns inspected. Therefore, there could be some misclassified particles in our sampling. To overcome this issue as much as possible, we consider regular particles that have reached a final value of $\log_{10}(\mathrm{FLI}/n) < -4.5$, and, as chaotic particles, we consider those that have reached a final value of $\log_{10}(\mathrm{FLI}/n) > -4.5$. This arbitrarily threshold strategy is based on the expected properties of the FLI dynamic indicator and considers a subset of initial conditions that have already manifested clear regular or chaotic behaviour at \num{1e5} turns, while excluding those that do not yet have a clear classification.

The sets are then used to inspect the time evolution of both $\mathrm{FLI}/n$ and $\mathrm{FLI}^{\mathrm{WB}}$, to assess possible classification improvements in the latter compared to the first. These improvements can be, for example, an increased convergence rate in the indicator value or an increased spread between the values of regular and chaotic initial conditions.

\begin{figure}[htp]
    \centering
    \includegraphics[width=1.0\textwidth]{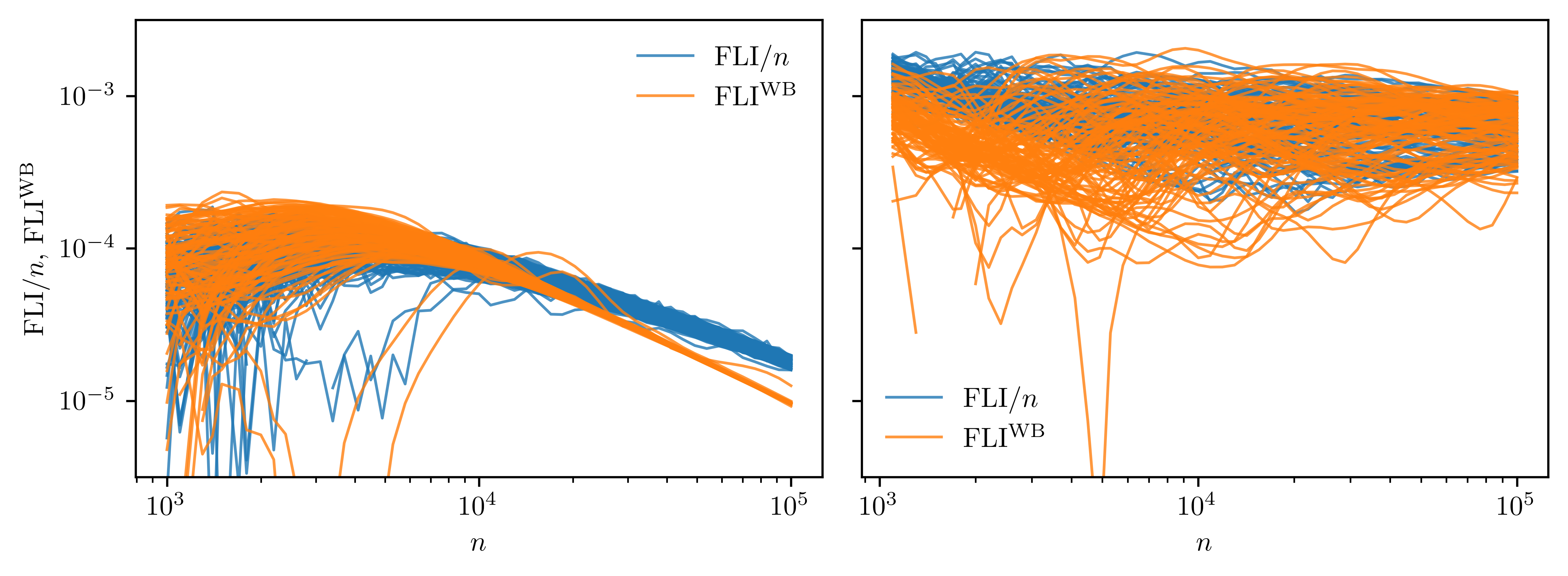}
    \caption{Time evolution of FLI computed using either a standard mean ($\mathrm{FLI}/n$) or the Birkhoff averaging ($\mathrm{FLI}^{\mathrm{WB}}$). Left plot: indicators computed for a set of $100$ regular initial conditions, above $10^4$ turns, a slightly faster convergence to zero is observed for $\mathrm{FLI}^{\mathrm{WB}}$. Right plot: indicators computed for a set of $100$ chaotic initial conditions. A slight difference in convergence rate is observed for low $n$ values, before reaching a saturation value of the indicator of the order of $10^{-3}-10^{-3.5}$. The results are obtained using the worst seed and $\zeta_0=$\SI{0.0761}{\meter}.)}
    \label{fig:fli_compare_mean_birk_2}
\end{figure}

In Fig.~\ref{fig:fli_compare_mean_birk_2} (left), the comparison between the two indicators is made for a subset of the set of regular initial conditions. It is possible to observe how, for regular initial conditions, Birkhoff averaging does slightly improve the convergence rate, as the indicator tends to the zero value. This improvement can be clearly observed above the sampling of $n=10^4$ turns, while before that value the two indicators appear to have similar strong fluctuations around the range of values $[-3.75, -5.0]$. This initial behaviour might be related to the low number of turns considered, along with the renormalisation time interval $\tau=10^2$, which finally causes the number of samples to be less than $100$ for both averaging methods, possibly leading to a delay in the appearance of the expected convergence rate.

In Fig.~\ref{fig:fli_compare_mean_birk_2} (right), we show the comparison for the subset of chaotic initial conditions. In this case, a saturation region is observed for the indicator value of the order of $10^{-3}-10^{-3.5}$ for both indicators. When this value is reached, both indicators oscillate around it. However, the slope with which this non-zero value is reached is different for the two indicators and is higher in absolute value for $\mathrm{FLI}/n$ than for $\mathrm{FLI}^{\mathrm{WB}}$. This, combined with the fact that both indicators showed a similar convergence rate for regular initial conditions, suggests a greater difference in the convergence rate of regular and chaotic initial conditions for $\mathrm{FLI}^{\mathrm{WB}}$, compared to $\mathrm{FLI}/n$. We must also point out how some chaotic initial conditions exhibit large fluctuations in $\mathrm{FLI}^{\mathrm{WB}}$ before the saturation point, reaching values comparable to the regular initial conditions. These isolated cases might be artefacts caused by the Birkhoff weights, which amplify certain modes in the time series when the chaotic behaviour still has not fully manifested.

\begin{figure}[htb]
    \centering
    \includegraphics[width=0.75\textwidth]{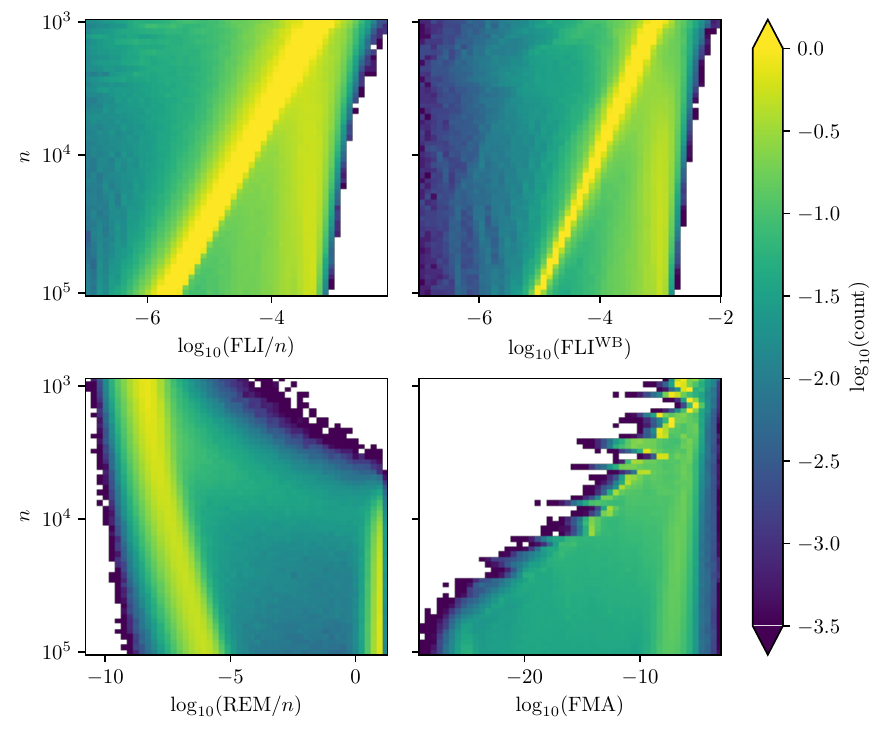}
    \caption{Distribution of values of the various dynamic indicators as a function of time for a realistic HL-LHC lattice. For low values of the number of turns $n$, the distribution is in general represented by a uni-modal function. For higher values of $n$, we can see the formation of either two separate clusters, making the distribution bi-modal, or an individual cluster with a significant tail. $\log_{10}(FMA)$ constitutes an exception, as it evolves forming a tri-modal distribution. (HL-LHC lattice used: worst seed, $\zeta_0=$\SI{0.0761}{\meter}.)}
    \label{fig:overview2}
\end{figure}

\begin{figure}[htb]
    \centering
    \includegraphics[width=1.0\textwidth]{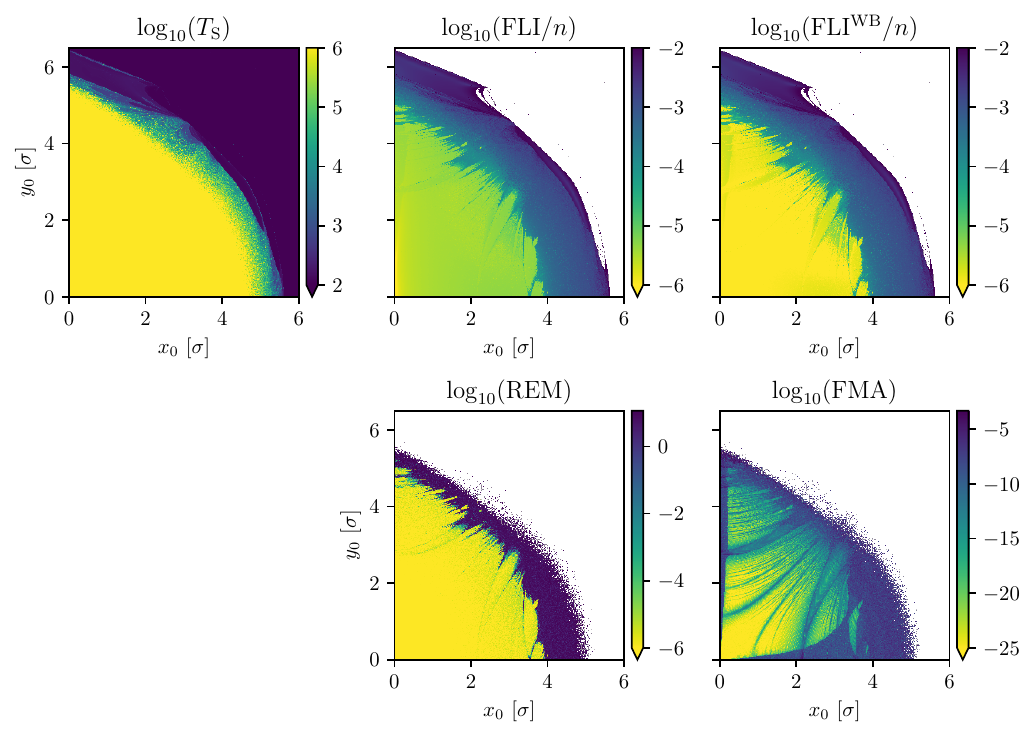}
    \caption{Colour maps of the various dynamic indicators for a realistic HL-LHC lattice, using the worst seed and $\zeta_0=\SI{0.0761}{\meter}$, evaluated at $n=\num{1e5}$ turns. It can be seen how the indicators globally highlight the same structures in phase space, except for FMA, which also shows additional structures.}
    \label{fig:overview}
\end{figure}

Despite these isolated oscillations, the improvement brought about by the Birkhoff averages can be appreciated by comparing the evolution of the value distribution of $\log_{10}(\mathrm{FLI}/n)$ and $\log_{10}(FLI^\mathrm{WB})$ reported in Fig.~\ref{fig:overview2}, since there the time evolution of the value distribution is shown for all the initial conditions shown in Fig.~\ref{fig:overview}. The part of the distribution corresponding to the regular initial conditions reaches its peak (yellow band) and moves toward zero with increasing $n$. However, the displacement towards zero is faster for $\mathrm{FLI}^{\mathrm{WB}}$, and the peak is also narrower, potentially allowing for a better binary classification of regular and chaotic orbits, as pointed out in~\cite{PhysRevE.107.064209}.

\subsection{Dependence of FMA from the longitudinal dynamics}

As we can observe in Fig.~\ref{fig:overview}, the chaotic structures highlighted by FMA are very different from those highlighted by the other dynamic indicators. This is because FMA is generally sensitive to tune changes, which are not necessarily related to chaotic dynamics but rather related to the presence of resonances or tune modulation.

Another feature emerges when considering the evolution of the distribution of indicator values as a function of time. In fact, if we compare the evolution of the value distribution of FMA, we can observe, in Fig.~\ref{fig:overview2}, how FMA, unlike the other dynamic indicators, does not show a tendency to create a sharp bi-modal distribution.

\begin{figure}[ht]
    \centering
    \includegraphics[width=1.0\textwidth]{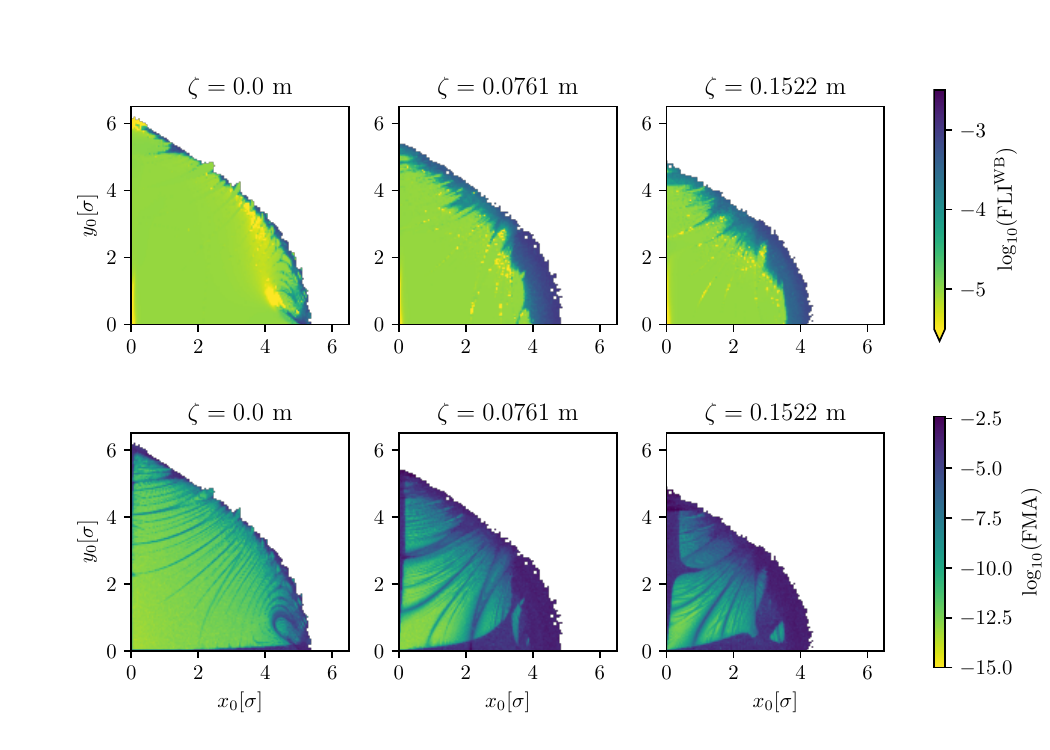}
    \caption{$\log_{10}(\mathrm{FMA})$ (top row) and $\log_{10}(\mathrm{FLI}^{\mathrm{WB}}(\hat{x}))$ (bottom row) both evaluated on the same seed on the worst seed and for the three values of $\zeta_0$ at $n=\num{1e5}$ turns. The differences between the two indicators are enhanced for larger values of $\zeta_0$.}
    \label{fig:fma_vs_fli}
\end{figure}

To further highlight this characteristic of FMA, we can observe how the indicator is particularly sensitive to the presence of longitudinal dynamics. Indeed, the longitudinal dynamics couples with the transverse one, also introducing tune modulation via a finite value of the chromaticity. In Fig.~\ref{fig:fma_vs_fli}, we present the FMA evaluated for the worst seed for the three values of $\zeta_0$ inspected, and we compare the resulting structures with those highlighted by $\mathrm{FLI}^{\mathrm{WB}}$. Two essential features can be observed. The first is that the chaotic regions at the border of the stable region resemble very much the two indicators, which show that their behaviour is similar. The second is a strong difference between the two indicators in the region close to the vertical and horizontal axis. This difference grows significantly as a function of $\zeta_0$. 
With increasing values of $\zeta_0$, the chaotic region detected by the FMA extends toward the origin along the vertical axis. Furthermore, the width of this chaotic region increases with $\zeta_0$. It is quite clear that the chaotic behaviour detected by FMA is an artefact related to the presence of a strong modulation of the tunes. In this sense, this observation suggests that the use of FMA to identify chaotic regions in phase space is taken with a grain of salt whenever modulation of linear tunes is present.   
\subsection{Overview of REM behaviour}
In the performance review presented in~\cite{PhysRevE.107.064209}, REM proved to be the most effective dynamic indicator to quickly achieve a binary distinction between regular and chaotic initial conditions. This is mainly due to the different scaling law followed by the indicator for regular and chaotic initial conditions, which leads to a sharper bi-modal distribution of the indicator values. 

When applying REM to the HL-LHC lattices, we observe with high clarity the chaotic regions that are present in the phase space. In Fig.~\ref{fig:overview}, we can qualitatively compare the results obtained by REM, calculated for an example seed, with those obtained with $\mathrm{FLI}^{\mathrm{WB}}$. We can observe how REM tends to highlight chaotic regions qualitatively sharper than $\mathrm{FLI}^{\mathrm{WB}}$. However, we stress that, differently from previous research based on simpler accelerator models, it is not possible to establish a ground truth for the chaotic regions in the realistic HL-LHC lattice. However, we expect, based on the results on simpler models, that REM is able to better highlight chaotic regions in the phase space. 

The behaviour of the distribution of the indicator values can be appreciated by comparing the evolution of the value distribution of $\log_{10}($REM$)$ and $\log_{10}(\mathrm{FLI}^\mathrm{WB})$ as a function of time, as shown in Fig.~\ref{fig:overview2}. In fact, $\log_{10}($REM$)$ tends to a bi-modal distribution much faster than $\log_{10}(\mathrm{FLI}^{\mathrm{WB}})$ and in such a way that a threshold for detecting chaotic initial conditions would be almost independent of $n$.

\begin{figure}
    \centering
    \includegraphics[width=\textwidth]{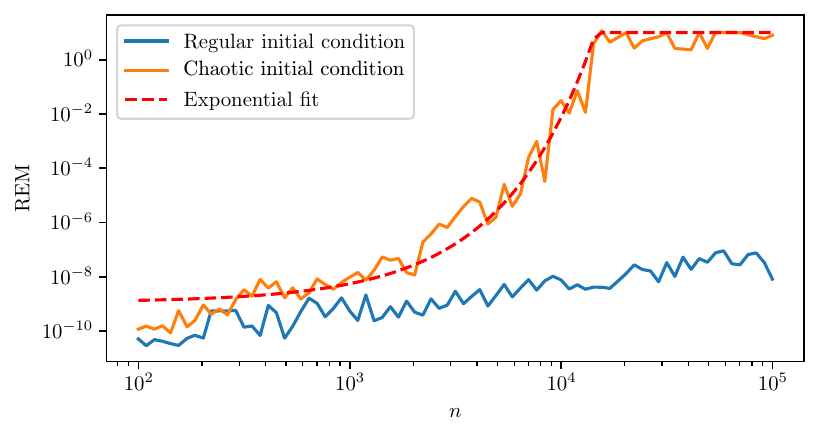}
    \caption{Evolution of REM for a regular (blue) and a chaotic (orange) initial condition. The case of the regular initial condition shows a power-law behaviour, while that of the chaotic initial condition shows an exponential-like behaviour with a saturation value of the order of \num{1e1}. An exponential fit of the chaotic initial condition to reconstruct the value of the MLE, is also presented.}
    \label{fig:rem_evolution}
\end{figure}

A more detailed overview of the behaviour of REM for different initial conditions is presented in Fig.~\ref{fig:rem_evolution}, where a comparison is presented between the power-law evolution of REM of a regular initial condition and the exponential evolution of a chaotic one. Note how REM, for the chaotic initial condition, saturates to values up to $~10^1$. This saturation value is comparable to the diameter of the explorable phase space, and it indicates that the chaotic behaviour eventually led the backtracked particle to completely lose track of the initial path, exploring completely different regions of the phase space. 

\begin{figure}
    \centering
    \includegraphics[width=\textwidth]{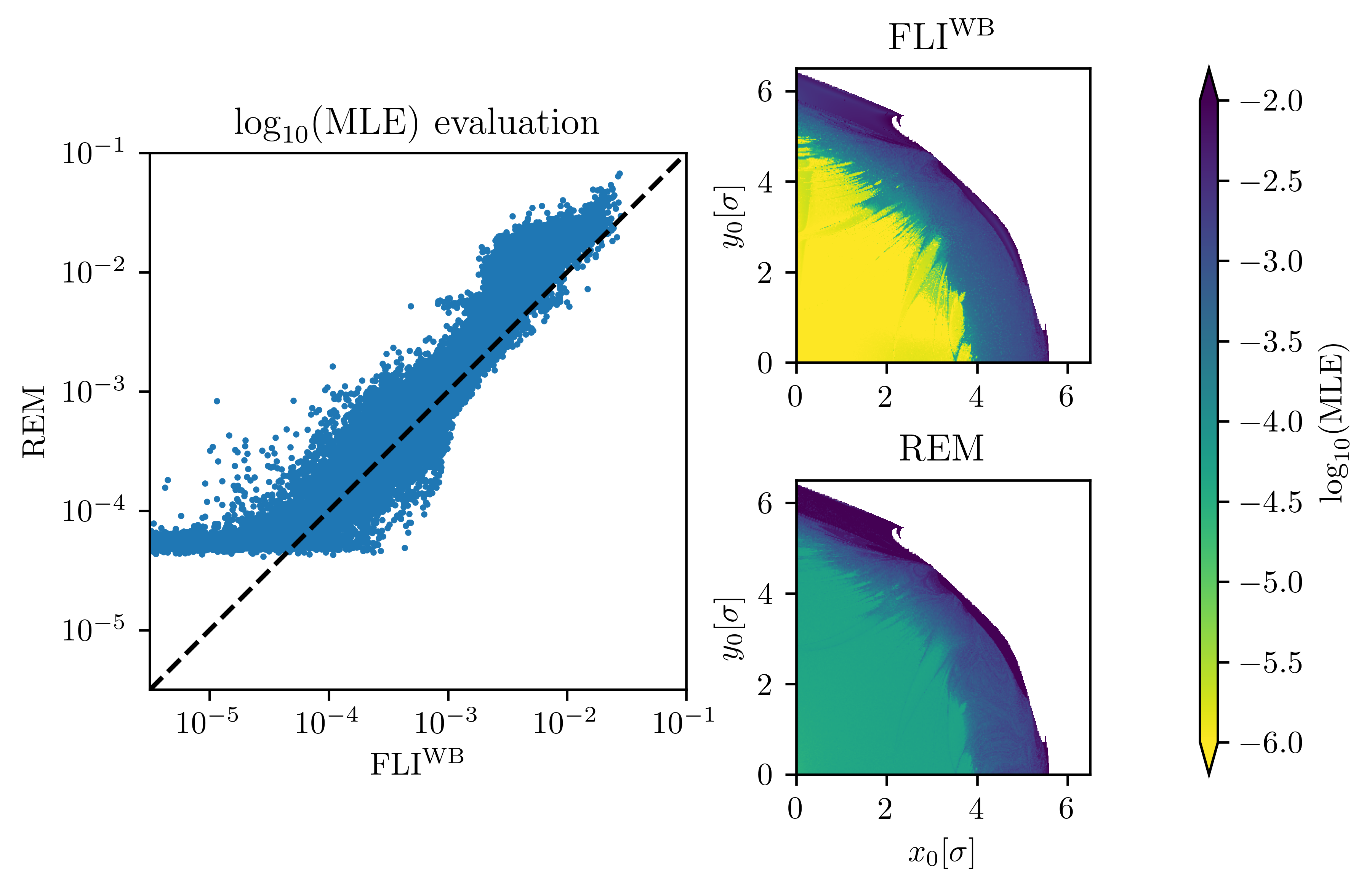}
    \caption{Right: value of MLE for the Cartesian grid of initial conditions reconstructed using REM and $\mathrm{FLI}^{\mathrm{WB}}$ for the case of the worst seed. Left: correlation plot of MLE reconstructed using REM and $\mathrm{FLI}^{\mathrm{WB}}$.}
    \label{fig:rem_vs_fli}
\end{figure}

By fitting an exponential law to the evolution of REM, while also excluding saturated parts of the data, it is possible to recover the value of the MLE governing the timescale of the process. We present the value obtained for all initial conditions and compare it to the value obtained from $\mathrm{FLI}^{\mathrm{WB}}$ in Fig.~\ref{fig:rem_vs_fli}. We can see how the two indicators agree well, except for a few outliers, which might be related to cases in which the chaotic behaviour has not yet fully manifested for the $\mathrm{FLI}^{\mathrm{WB}}$ indicator to be able to correctly evaluate the MLE.

The promising performance of REM in highlighting chaotic structures in the phase space, along with its straightforward numerical implementation, makes it a very interesting tool for studying phase-space structures in realistic accelerator lattices.

\clearpage
\bibliographystyle{unsrt}
\bibliography{bibliography}

\begin{thebibliography}{10}

\bibitem{PANICHI201653}
F.~Panichi, L.~Ciotti, and G.~Turchetti.
\newblock {Fidelity and reversibility in the restricted three body problem}.
\newblock {\em Communications in Nonlinear Science and Numerical Simulation},
  35:53--68, 2016.

\bibitem{Panichi2017}
{F. Panichi, K. Go{\'z}dziewski and G. Turchetti}.
\newblock {The reversibility error method (REM): a new, dynamical fast
  indicator for planetary dynamics}.
\newblock {\em MNRAS}, 468(1):469--491, 6 2017.

\bibitem{Lega2016fli}
E.~Lega, M.~Guzzo, and C.~Froeschl{\'e}.
\newblock {Theory and Applications of the Fast Lyapunov Indicator (FLI)
  Method}.
\newblock In C.~Skokos, G.~A. Gottwald, and J.~Laskar, editors, {\em Chaos
  Detection and Predictability}, pages 35--54, Berlin, Heidelberg, 2016.
  Springer Berlin Heidelberg.

\bibitem{Guzzo2023}
M.~Guzzo and E.~Lega.
\newblock Theory and applications of fast lyapunov indicators to model problems
  of celestial mechanics.
\newblock {\em Celestial Mech. Dyn. Astron.}, 135(4):37, 2023.

\bibitem{dynap1}
M.~Giovannozzi, W.~Scandale, and E.~Todesco.
\newblock {Prediction of long-term stability in large hadron colliders}.
\newblock {\em Part. Accel.}, 56:195, 1997.

\bibitem{PhysRevAccelBeams.23.084601}
K.~Hwang, C.~Mitchell, and R.~Ryne.
\newblock Rapidly converging chaos indicator for studying dynamic aperture in a
  storage ring with space charge.
\newblock {\em Phys. Rev. Accel. Beams}, 23:084601, 8 2020.

\bibitem{LI2021164936}
Y.~Li, Y.~Hao, K.~Hwang, R.~Rainer, A.~He, and A.~Liu.
\newblock {Fast dynamic aperture optimization with forward-reversal
  integration}.
\newblock {\em Nucl. Instrum. Methods Phys. Res., Sect. A}, 988:164936, 2021.

\bibitem{montanari:ipac2022-mopost042}
C.E. Montanari, A.~Bazzani, M.~Giovannozzi, and G.~Turchetti.
\newblock {Using Dynamic Indicators for Probing Single-Particle Stability in
  Circular Accelerators}.
\newblock In {\em Proc. IPAC'22}, number~13 in International Particle
  Accelerator Conference, pages 168--171. JACoW Publishing, Geneva,
  Switzerland, 07 2022.

\bibitem{PhysRevE.53.4067}
E.~Todesco and M.~Giovannozzi.
\newblock {Dynamic aperture estimates and phase-space distortions in nonlinear
  betatron motion}.
\newblock {\em Phys. Rev. E}, 53(4):4067--4076, 4 1996.

\bibitem{invlog}
M.~Giovannozzi, W.~Scandale, and E.~Todesco.
\newblock {Dynamic aperture extrapolation in presence of tune modulation}.
\newblock {\em Phys. Rev.}, E57(3):3432, 3 1998.

\bibitem{Bazzani:2019csk}
A.~Bazzani, M.~Giovannozzi, E.~H. Maclean, C.~E. Montanari, F.~F. Van~der
  Veken, and W.~Van~Goethem.
\newblock {Advances on the modeling of the time evolution of dynamic aperture
  of hadron circular accelerators}.
\newblock {\em Phys. Rev. Accel. Beams}, 22:104003, 10 2019.

\bibitem{Bruning:782076}
O.~S. Br\"uning, P.~Collier, Ph. Lebrun, S.~Myers, R.~Ostojic, J.~Poole, and
  P.~Proudlock.
\newblock {\em {LHC Design Report}}.
\newblock {CERN} Yellow Reports: Monographs. CERN, Geneva, 2004.

\bibitem{BejarAlonso:2749422}
O.~Aberle, I~Béjar~Alonso, O~Brüning, P~Fessia, L~Rossi, L~Tavian,
  M~Zerlauth, C.~Adorisio, A.~Adraktas, M.~Ady, J.~Albertone, L.~Alberty,
  M.~Alcaide~Leon, A.~Alekou, D.~Alesini, B.~Almeida Ferreira, P.~Alvarez
  Lopez, G.~Ambrosio, P.~Andreu~Munoz, M.~Anerella, D.~Angal-Kalinin,
  F.~Antoniou, G.~Apollinari, A.~Apollonio, R.~Appleby, G.~Arduini, B.~Arias
  Alonso, K.~Artoos, S.~Atieh, B.~Auchmann, V.~Badin, T.~Baer, D.~Baffari,
  V.~Baglin, M.~Bajko, A.~Ball, A.~Ballarino, S.~Bally, T.~Bampton, D.~Banfi,
  R.~Barlow, M.~Barnes, J.~Barranco, L.~Barthelemy, W.~Bartmann, H.~Bartosik,
  E.~Barzi, M.~Battistin, P.~Baudrenghien, I.~Bejar Alonso, S.~Belomestnykh,
  A.~Benoit, I.~Ben-Zvi, A.~Bertarelli, S.~Bertolasi, C.~Bertone, B.~Bertran,
  P.~Bestmann, N.~Biancacci, A.~Bignami, N.~Bliss, C.~Boccard, Y.~Body,
  J.~Borburgh, B.~Bordini, F.~Borralho, R.~Bossert, L.~Bottura, A.~Boucherie,
  R.~Bozzi, C.~Bracco, E.~Bravin, G.~Bregliozzi, D.~Brett, A.~Broche,
  K.~Brodzinski, F.~Broggi, R.~Bruce, M.~Brugger, O.~Brüning, X.~Buffat,
  H.~Burkhardt, J.~Burnet, A.~Burov, G.~Burt, R.~Cabezas, Y.~Cai, R.~Calaga,
  S.~Calatroni, O.~Capatina, T.~Capelli, P.~Cardon, E.~Carlier, F.~Carra,
  A.~Carvalho, L.R. Carver, F.~Caspers, G.~Cattenoz, F.~Cerutti, A.~Chancé,
  M.~Chastre Rodrigues, S.~Chemli, D.~Cheng, P.~Chiggiato, G.~Chlachidze,
  S.~Claudet, JM. Coello De~Portugal, C.~Collazos, J.~Corso, S.~Costa~Machado,
  P.~Costa~Pinto, E.~Coulinge, M.~Crouch, P.~Cruikshank, E.~Cruz~Alaniz,
  M.~Czech, K.~Dahlerup-Petersen, B.~Dalena, G.~Daniluk, S.~Danzeca, H.~Day,
  J.~De~Carvalho~Saraiva, D.~De~Luca, R.~De~Maria, G.~De~Rijk, S.~De~Silva,
  B.~Dehning, J.~Delayen, Q.~Deliege, B.~Delille, F.~Delsaux, R.~Denz,
  A.~Devred, A.~Dexter, B.~Di~Girolamo, D.~Dietderich, J.W. Dilly, A.~Doherty,
  N.~Dos~Santos, A.~Drago, D.Drskovic, D.~Duarte Ramos, L.~Ducimetière,
  I.~Efthymiopoulos, K.~Einsweiler, L.~Esposito, J.~Esteban~Muller, S.~Evrard,
  P.~Fabbricatore, S.~Farinon, S.~Fartoukh, A.~Faus-Golfe, G.~Favre, H.~Felice,
  B.~Feral, G.~Ferlin, P.~Ferracin, A.~Ferrari, L.~Ferreira, P.~Fessia,
  L.~Ficcadenti, S.~Fiotakis, L.~Fiscarelli, M.~Fitterer, J.~Fleiter,
  G.~Foffano, E.~Fol, R.~Folch, K.~Foraz, A.~Foussat, M.~Frankl, O.~Frasciello,
  M.~Fraser, P.~Freijedo Menendez, J-F. Fuchs, S.~Furuseth, A.~Gaddi,
  M.~Gallilee, A.~Gallo, R.~Garcia Alia, H.~Garcia Gavela, J.~Garcia Matos,
  H.~Garcia~Morales, A.~Garcia-Tabares Valdivieso, C.~Garino, C.~Garion,
  J.~Gascon, Ch. Gasnier, L.~Gentini, C.~Gentsos, A.~Ghosh, L.~Giacomel,
  K.~Gibran Hernandez, S.~Gibson, C.~Ginburg, F.~Giordano, M.~Giovannozzi,
  B.~Goddard, P.~Gomes, M.~Gonzalez De La Aleja~Cabana, P.~Goudket, E.~Gousiou,
  P.~Gradassi, A.~Granadeiro Costa, L.~Grand-Clément, S.~Grillot, JC.
  Guillaume, M.~Guinchard, P.~Hagen, T.~Hakulinen, B.~Hall, J.~Hansen,
  N.~Heredia~Garcia, W.~Herr, A.~Herty, C.~Hill, M.~Hofer, W.~Höfle,
  B.~Holzer, S.~Hopkins, J.~Hrivnak, G.~Iadarola, A.~Infantino, S.~Izquierdo
  Bermudez, S.~Jakobsen, M.A. Jebramcik, B.~Jenninger, E.~Jensen, M.~Jones,
  R.~Jones, T.~Jones, J.~Jowett, M.~Juchno, C.~Julie, T.~Junginger, V.~Kain,
  D.~Kaltchev, N.~Karastathis, P.~Kardasopoulos, M.~Karppinen, J.~Keintzel,
  R.~Kersevan, F.~Killing, G.~Kirby, M.~Korostelev, N.~Kos, S.~Kostoglou,
  I.~Kozsar, A.~Krasnov, S.~Krave, L.~Krzempek, N.~Kuder, A.~Kurtulus,
  R.~Kwee-Hinzmann, F.~Lackner, M.~Lamont, A.L. Lamure, L.~Lari m,
  M.~Lazzaroni, M.~Le~Garrec, A.~Lechner, T.~Lefevre, R.~Leuxe, K.~Li, Z.~Li,
  R.~Lindner, B.~Lindstrom, C.~Lingwood, C.~Löffler, C.~Lopez, LA.
  Lopez-Hernandez, R.~Losito, F.~Maciariello, P.~Macintosh, E.H. Maclean,
  A.~Macpherson, P.~Maesen, C.~Magnier, H.~Mainaud Durand, L.~Malina,
  M.~Manfredi, F.~Marcellini, M.~Marchevsky, S.~Maridor, G.~Marinaro,
  K.~Marinov, T.~Markiewicz, A.~Marsili, P.~Martinez~Urioz, M.~Martino,
  A.~Masi, T.~Mastoridis, P.~Mattelaer, A.~May, J.~Mazet, S.~Mcilwraith,
  E.~McIntosh, L.~Medina~Medrano, A.~Mejica~Rodriguez, M.~Mendes, P.~Menendez,
  M.~Mensi, A.~Mereghetti, D.~Mergelkuhl, T.~Mertens, L.~Mether, E.~Métral,
  M.~Migliorati, A.~Milanese, P.~Minginette, D.~Missiaen, T.~Mitsuhashi,
  M.~Modena, N.~Mokhov, J.~Molson, E.~Monneret, E.~Montesinos,
  R.~Moron-Ballester, M.~Morrone, A.~Mostacci, N.~Mounet, P.~Moyret, P.~Muffat,
  B.~Muratori, Y.~Muttoni, T.~Nakamoto, M.~Navarro-Tapia, H.~Neupert, L.~Nevay,
  T.~Nicol, E.~Nilsson, P.~Ninin, A.~Nobrega, C.~Noels, E.~Nolan, Y.~Nosochkov,
  FX. Nuiry, L.~Oberli, T.~Ogitsu, K.~Ohmi, Olave R., J.~Oliveira, Ph. Orlandi,
  P.~Ortega, J.~Osborne, T.~Otto, L.~Palumbo, S.~Papadopoulou,
  Y.~Papaphilippou, K.~Paraschou, C.~Parente, S.~Paret, H.~Park, V.~Parma, Ch.
  Pasquino, A.~Patapenka, L.~Patnaik, S.~Pattalwar, J.~Payet, G.~Pechaud,
  D.~Pellegrini, P.~Pepinster, J.~Perez, J.~Perez Espinos, A.~Perillo Marcone,
  A.~Perin, P.~Perini, T.H.B. Persson, T.~Peterson, T.~Pieloni, G.~Pigny, J.P.
  Pinheiro~de Sousa, O.~Pirotte, F.~Plassard, M.~Pojer, L.~Pontercorvo,
  A.~Poyet, D.~Prelipcean, H.~Prin, R.~Principe, T.~Pugnat, J.~Qiang,
  E.~Quaranta, H.~Rafique, I.~Rakhno, D.~Ramos Duarte, A.~Ratti, E.~Ravaioli,
  M.~Raymond, S.~Redaelli, T.~Renaglia, D.~Ricci, G.~Riddone, J.~Rifflet,
  E.~Rigutto, T.~Rijoff, R.~Rinaldesi, O.~Riu~Martinez, L.~Rivkin,
  F.~Rodriguez~Mateos, S.~Roesler, I.~Romera~Ramirez, A.~Rossi, L.~Rossi,
  V.~Rude, G.~Rumolo, J.~Rutkovksi, M.~Sabate~Gilarte, G.~Sabbi, T.~Sahner,
  R.~Salemme, B.~Salvant, F.~Sanchez Galan, A.~Santamaria~Garcia,
  I.~Santillana, C.~Santini, O.~Santos, P.~Santos Diaz, K.~Sasaki, F.~Savary,
  A.~Sbrizzi, M.~Schaumann, C.~Scheuerlein, J.~Schmalzle, H.~Schmickler,
  R.~Schmidt, D.~Schoerling, M.~Segreti, M.~Serluca, J.~Serrano, J.~Sestak,
  E.~Shaposhnikova, D.~Shatilov, A.~Siemko, M.~Sisti, M.~Sitko, J.~Skarita,
  E.~Skordis, K.~Skoufaris, G.~Skripka, D.~Smekens, Z.~Sobiech, M.~Sosin,
  M.~Sorbio, F.~Soubelet, B.~Spataro, G.~Spiezia, G.~Stancari, M.~Staterao,
  J.~Steckert, G.~Steele, G.~Sterbini, M.~Struik, M.~Sugano, A.~Szeberenyi,
  M.~Taborelli, C.~Tambasco, R.~Tavares Rego, L.~Tavian, B.~Teissandier,
  N.~Templeton, M.~Therasse, H.~Thiesen, E.~Thomas, A.~Toader, E.~Todesco,
  R.~Tomás, F.~Toral, R.~Torres-Sanchez, G.~Trad, N.~Triantafyllou, I.~Tropin,
  A.~Tsinganis, J.~Tuckamantel, J.~Uythoven, A.~Valishev, F.~Van Der~Veken,
  R.~Van~Weelderen, A.~Vande~Craen, B.~Vazquez De~Prada, F.~Velotti,
  S.~Verdu~Andres, A.~Verweij, N.~Vittal Shetty, V.~Vlachoudis, G.~Volpini,
  U.~Wagner, P.~Wanderer, M.~Wang, X.~Wang, R.~Wanzenberg, A.~Wegscheider,
  S.~Weisz, C.~Welsch, M.~Wendt, J.~Wenninger, W.~Weterings, S.~White,
  K.~Widuch, A.~Will, G.~Willering, D.~Wollmann, A.~Wolski, J.~Wozniak, Q.~Wu,
  B.~Xiao, L.~Xiao, Q.~Xu, Y.~Yakovlev, S.~Yammine, Y.~Yang, M.~Yu,
  I.~Zacharov, O.~Zagorodnova, C.~Zannini, C.~Zanoni, M.~Zerlauth,
  F.~Zimmermann, A.~Zlobin, M.~Zobov, and I.~Zurbano~Fernandez.
\newblock {\em {High-Luminosity Large Hadron Collider (HL-LHC): Technical
  design report}}.
\newblock CERN Yellow Reports: Monographs. CERN, Geneva, 2020.

\bibitem{Arduini_2016}
G.~Arduini, J.~Barranco, A.~Bertarelli, N.~Biancacci, R.~Bruce, O.~Br\"{u}ning,
  X.~Buffat, Y.~Cai, L.R. Carver, S.~Fartoukh, M.~Giovannozzi, G.~Iadarola,
  K.~Li, A.~Lechner, L.~Medina Medrano, E.~M\'{e}tral, Y.~Nosochkov,
  Y.~Papaphilippou, D.~Pellegrini, T.~Pieloni, J.~Qiang, S.~Redaelli,
  A.~Romano, L.~Rossi, G.~Rumolo, B.~Salvant, M.~Schenk, C.~Tambasco,
  R.~Tom\'{a}s, S.~Valishev, and F.F.~Van der Veken.
\newblock {High Luminosity LHC: challenges and plans}.
\newblock {\em Journal of Instrumentation}, 11(12):C12081, 12 2016.

\bibitem{Das_2017}
S.~Das, Y.~Saiki, E.~Sander, and J.~A. Yorke.
\newblock {Quantitative quasiperiodicity}.
\newblock {\em Nonlinearity}, 30(11):4111--4140, 10 2017.

\bibitem{Das_2018}
Suddhasattwa Das and James~A Yorke.
\newblock {Super convergence of ergodic averages for quasiperiodic orbits}.
\newblock {\em Nonlinearity}, 31(2):491--501, 1 2018.

\bibitem{Panichi2016}
{F. Panichi, L. Ciotti and G. Turchetti}.
\newblock {Fidelity and reversibility in the restricted three body problem}.
\newblock {\em Commun. Nonlinear Sci. Numer. Simul.}, 35:53--68, 2016.

\bibitem{Laskar1999}
J.~Laskar.
\newblock {Introduction to Frequency Map Analysis}.
\newblock In C.~Sim\'{o}, editor, {\em Hamiltonian Systems with Three or More
  Degrees of Freedom}, pages 134 --- 150, Dordrecht, 1999. Springer, Springer.

\bibitem{Laskar2003}
J.~Laskar.
\newblock {Frequency map analysis and quasiperiodic decompositions}.
\newblock {\em arXiv: Dynamical Systems}, 2003.

\bibitem{laskar1995frequency}
J.~Laskar.
\newblock {Frequency map analysis of an Hamiltonian system}.
\newblock In {\em AIP conference proceedings}, volume 344, pages 130--159.
  American Institute of Physics, 1995.

\bibitem{lega1996numerical}
E.~Lega and C.~Froeschl{\'e}.
\newblock {Numerical investigations of the structure around an invariant KAM
  torus using the frequency map analysis}.
\newblock {\em Physica D}, 95(2):97--106, 1996.

\bibitem{papaphilippou1996frequency}
Y.~Papaphilippou and J.~Laskar.
\newblock {Frequency map analysis and global dynamics in a galactic potential
  with two degrees of freedom.}
\newblock {\em Astron. Astrophys.}, 307:427--449, 1996.

\bibitem{papaphilippou1998global}
Y.~Papaphilippou and J.~Laskar.
\newblock {Global dynamics of triaxial galactic models through frequency map
  analysis}.
\newblock {\em Astron. Astrophys.}, 329:451--481, 1998.

\bibitem{Papaphilippou1999}
Y.~Papaphilippou.
\newblock {Global Dynamics of a Galactic Potential via Frequency Map Analysis}.
\newblock In C.~Sim{\'o}, editor, {\em {Hamiltonian Systems with Three or More
  Degrees of Freedom}}, pages 523--527, Dordrecht, 1999. Springer Netherlands.

\bibitem{laskar2000application}
J.~Laskar.
\newblock {Application of frequency map analysis}.
\newblock In {\em The Chaotic Universe: Proceedings of the Second ICRA Network
  Workshop, Rome, Pescara, Italy, 1-5 February 1999}, volume~10, page 115.
  World Scientific, 2000.

\bibitem{PhysRevSTAB.4.124201}
M.~Comunian, A.~Pisent, A.~Bazzani, G.~Turchetti, and S.~Rambaldi.
\newblock {Frequency map analysis of a three-dimensional particle in the core
  model of a high intensity linac}.
\newblock {\em Phys. Rev. ST Accel. Beams}, 4:124201, 12 2001.

\bibitem{1288929}
J.~Laskar.
\newblock {Frequency map analysis and particle accelerators}.
\newblock In {\em Proceedings of the 2003 Particle Accelerator Conference},
  volume~1, pages 378--382 Vol.1, 2003.

\bibitem{Papaphilippou:PAC03-RPPG007}
Y.~Papaphilippou, L.~Farvacque, J.~Laskar, and A.~Ropert.
\newblock {Probing the Non-Linear Dynamics of the ESRF Storage Ring with
  Experimental Frequency Maps}.
\newblock In {\em Proceedings of the 2003 Particle Accelerator Conference},
  volume~1, pages 3189--3191. IEEE, 2003.

\bibitem{PhysRevSTAB.6.114801}
L.~Nadolski and J.~Laskar.
\newblock {Review of single particle dynamics for third generation light
  sources through frequency map analysis}.
\newblock {\em Phys. Rev. ST Accel. Beams}, 6:114801, 11 2003.

\bibitem{shun2009nonlinear}
T.~Shun-Qiang, L.~Gui-Min, L.~Hao-Hu, H.~Jie, C.~Guang-Ling, and W.~Cheng-Lan.
\newblock {Nonlinear optimization of the modern synchrotron radiation storage
  ring based on frequency map analysis}.
\newblock {\em Chinese Physics C}, 33(2):127, 2009.

\bibitem{PhysRevSTAB.14.014001}
D.~Shatilov, E.~Levichev, E.~Simonov, and M.~Zobov.
\newblock {Application of frequency map analysis to beam-beam effects study in
  crab waist collision scheme}.
\newblock {\em Phys. Rev. ST Accel. Beams}, 14:014001, 1 2011.

\bibitem{papaphilippou2014}
Y.~Papaphilippou.
\newblock {Detecting chaos in particle accelerators through the frequency map
  analysis method}.
\newblock {\em Chaos}, 24(2):024412, 2014.

\bibitem{tydecks:ipac18-mopmf057}
T.~Tydecks et~al.
\newblock {FCC-ee Dynamic Aperture Studies and Frequency Map Analysis}.
\newblock In {\em Proc. IPAC'18}, pages 244--246. JACoW Publishing, Geneva,
  Switzerland, 2018.

\bibitem{PhysRevAccelBeams.22.071002}
P.~Zisopoulos, Y.~Papaphilippou, and J.~Laskar.
\newblock {Refined betatron tune measurements by mixing beam position data}.
\newblock {\em Phys. Rev. Accel. Beams}, 22:071002, 7 2019.

\bibitem{PhysRevE.107.064209}
A.~Bazzani, M.~Giovannozzi, C.~E. Montanari, and G.~Turchetti.
\newblock Performance analysis of indicators of chaos for nonlinear dynamical
  systems.
\newblock {\em Phys. Rev. E}, 107:064209, 6 2023.

\bibitem{Bazzani:262179}
A.~Bazzani, G.~Servizi, E.~Todesco, and G.~Turchetti.
\newblock {\em {A normal form approach to the theory of nonlinear betatronic
  motion}}.
\newblock {CERN} Yellow Reports: Monographs. {CERN}, Geneva, 1994.

\bibitem{ab_fb_gt_AIP}
A.~Bazzani, F.~Brini, and G.~Turchetti.
\newblock {Diffusion of the Adiabatic Invariant for Modulated Symplectic Maps}.
\newblock {\em AIP Conf. Proc. 395, 129}, 1997.

\bibitem{Bazzani9948}
A.~Bazzani, S.~Siboni, and G.~Turchetti.
\newblock {Diffusion in Hamiltonian systems with a small stochastic
  perturbation}.
\newblock {\em Physica D}, 76(1):8--21, 1994.

\bibitem{froeschle1999weak}
C.~Froeschl{\'e} and E.~Lega.
\newblock {Weak chaos and diffusion in Hamiltonian systems}.
\newblock In {\em The Dynamics of Small Bodies in the Solar System}, pages
  463--502. Springer, Berlin, Heidelberg, 1999.

\bibitem{Morbidelli1995}
A.~Morbidelli and C.~Froeschl{\'e}.
\newblock On the relationship between lyapunov times and macroscopic
  instability times.
\newblock {\em Celestial Mech. Dyn. Astron.}, 63(2):227--239, 6 1995.

\bibitem{morbidelli1995connection}
A.~Morbidelli and A.~Giorgilli.
\newblock {On a connection between KAM and Nekhoroshev's theorems}.
\newblock {\em Physica D}, 86(3):514--516, 1995.

\bibitem{CINCOTTA2022133101}
P.~M. Cincotta, C.~M. Giordano, and I.~I. Shevchenko.
\newblock Revisiting the relation between the lyapunov time and the instability
  time.
\newblock {\em Physica D}, 430:133101, 2022.

\bibitem{Giovannozzi:317866}
M.~Giovannozzi and E.~McIntosh.
\newblock {Development of parallel codes for the study of nonlinear beam
  dynamics}.
\newblock {\em Int. J. Mod. Phys. C}, 8:155--170, 11 1996.

\bibitem{DBLP:journals/corr/abs-1202-4347}
J.~Ghorpade, J.~Parande, M.~Kulkarni, and A.~Bawaskar.
\newblock {GPGPU Processing in CUDA Architecture}.
\newblock {\em CoRR}, abs/1202.4347, 2012.

\bibitem{iadarola2023xsuite}
G.~Iadarola, R.~De Maria, S.~Lopaciuk, A.~Abramov, X.~Buffat, D.~Demetriadou,
  L.~Deniau, P.~Hermes, P.~Kicsiny, P.~Kruyt, A.~Latina, L.~Mether,
  K.~Paraschou, Sterbini, F.~Van~Der Veken, P.~Belanger, P.~Niedermayer, D.~Di
  Croce, T.~Pieloni, and L.~Van Riesen-Haupt.
\newblock Xsuite: an integrated beam physics simulation framework, 2023.

\bibitem{xsuite}
G.~Iadarola.
\newblock {Xsuite documentation}, 2022.
\newblock https://xsuite.readthedocs.io.

\bibitem{De_Maria_2019}
R.~De Maria, J.~Andersson, V.K.~Berglyd Olsen, L.~Field, M.~Giovannozzi, P.D.
  Hermes, N.~H{\o}imyr, G.~Iadarola, S.~Kostoglou, E.~H. Maclean, E.~McIntosh,
  A.~Mereghetti, J.~Molson, D.~Pellegrini, T.~Persson, M.~Schwinzerl,
  B.~Dalena, T.~Pugnat, I.~Zacharov, and K.N. Sjobak.
\newblock {SixTrack Version 5: Status and New Developments}.
\newblock {\em Journal of Physics: Conference Series}, 1350(1):012129, 11 2019.

\bibitem{pang2014gpu}
X.~Pang and L.~Rybarcyk.
\newblock {GPU accelerated online multi-particle beam dynamics simulator for
  ion linear particle accelerators}.
\newblock {\em Comput. Phys. Commun.}, 185(3):744--753, 2014.

\bibitem{oeftiger:hb16-mopr025}
A.~Oeftiger and S.~Hegglin.
\newblock {Space Charge Modules for PyHEADTAIL}.
\newblock In {\em Proc. HB'16}, pages 124--129. JACoW Publishing, Geneva,
  Switzerland, 2016.

\bibitem{adelmann2019opal}
A.~Adelmann, P.~Calvo, M.~Frey, A.~Gsell, U.~Locans, C.~Metzger-Kraus,
  N.~Neveu, C.~Rogers, S.~Russell, S.~Sheehy, et~al.
\newblock {OPAL a versatile tool for charged particle accelerator simulations}.
\newblock {\em arXiv preprint arXiv:1905.06654}, 2019.

\bibitem{schwinzerl:ipac21-thpab190}
M.~Schwinzerl, H.~Bartosik, R.~De Maria, G.~Iadarola, A.~Oeftiger, and
  K.~Paraschou.
\newblock {Optimising and Extending a Single-Particle Tracking Library for High
  Parallel Performance}.
\newblock In {\em Proc. IPAC'21}, pages 4146--4149. JACoW Publishing, Geneva,
  Switzerland, 2021.

\bibitem{hermes:ipac2022-mopost045}
P.D. Hermes, R.~Bruce, R.~De Maria, M.~Giovannozzi, G.~Iadarola, D.~Mirarchi,
  and S.~Redaelli.
\newblock {A Novel Tool for Beam Dynamics Studies with Hollow Electron Lenses}.
\newblock In {\em Proc. IPAC'22}, number~13 in International Particle
  Accelerator Conference, pages 176--179. JACoW Publishing, Geneva,
  Switzerland, 07 2022.

\bibitem{iliakis2022enabling}
K.~Iliakis, H.~Timko, S.~Xydis, P.~Tsapatsaris, and D.~Soudris.
\newblock {Enabling Large Scale Simulations for Particle Accelerators}.
\newblock {\em IEEE Trans. Parallel Distrib. Syst.}, 33(12):4425--4439, 2022.

\bibitem{vanriesen-haupt:ipac2024-wepr05}
L.~van Riesen-Haupt~et al.
\newblock Benchmarking equilibrium emittance simulation tools for the future
  circular collider.
\newblock In {\em Proc. IPAC'24}, number~15 in IPAC'24 - 15th International
  Particle Accelerator Conference, pages 2461--2464. JACoW Publishing, Geneva,
  Switzerland, 05 2024.

\bibitem{fornara:ipac2024-thpc63}
A.~Fornara et~al.
\newblock Emittance growth studies due to crab cavity induced amplitude noise
  in the sps.
\newblock In {\em Proc. IPAC'24}, number~15 in IPAC'24 - 15th International
  Particle Accelerator Conference, pages 3163--3166. JACoW Publishing, Geneva,
  Switzerland, 05 2024.

\bibitem{Skokos2010b}
Ch. Skokos.
\newblock {The Lyapunov Characteristic Exponents and Their Computation}.
\newblock In J.-J. Souchay and R.~Dvorak, editors, {\em Dynamics of Small Solar
  System Bodies and Exoplanets}, pages 63--135, Berlin, Heidelberg, 2010.
  Springer Berlin Heidelberg.

\bibitem{Bartolini:292773}
R.~Bartolini, A.~Bazzani, M.~Giovannozzi, W.~Scandale, and E.~Todesco.
\newblock {Tune evaluation in simulations and experiments}.
\newblock {\em Part. Accel.}, 52:147--177. 29 p, 1995.

\bibitem{Bartolini:316949}
R.~Bartolini, M.~Giovannozzi, W.~Scandale, A.~Bazzani, and E.~Todesco.
\newblock {Precise measurement of the betatron tune}.
\newblock {\em Part. Accel.}, 55:1--10, 1996.

\bibitem{russo:ipac2021-thpab189}
G.~Russo, G.~Franchetti, and M.~Giovannozzi.
\newblock {New Techniques to Compute the Linear Tune}.
\newblock In {\em Proc. IPAC'21}, number~12 in International Particle
  Accelerator Conference, pages 4142--4145, Geneva, 08 2021. JACoW Publishing,
  Geneva, Switzerland.

\bibitem{hllhc15update}
R.~De~Maria.
\newblock {HL-LHC Optics v1.5 Update}, 7 2019.

\bibitem{hllhc15url}
R.~De~Maria.
\newblock {HL-LHC MAD-X optics V1.5 Tag v0.1}, 1 2020.

\bibitem{madx}
{MAD - Methodical Accelerator Design}.
\newblock \url{https://mad.web.cern.ch/mad/}.

\bibitem{Tancredi_2001}
G.~Tancredi, A.~Sánchez, and F.~Roig.
\newblock A comparison between methods to compute lyapunov exponents.
\newblock {\em Astron. J.}, 121(2):1171, 2 2001.

\bibitem{KAM1}
A.~N. Kolmogorov.
\newblock {On the Conservation of Conditionally Periodic Motions under Small
  Perturbation of the Hamiltonian}.
\newblock {\em Dokl. Akad. Nauk SSR}, 98:527, 1954.

\bibitem{KAM2}
J.~Moser.
\newblock {On invariant curves of area-preserving mappings of an annulus}.
\newblock {\em Nachr. Akad. Wiss. G\"{o}ttingen Math.-Phys. Kl.}, II:1, 1962.

\bibitem{KAM3}
V.I. Arnold.
\newblock {Proof of a theorem of A.N.~Kolmogorov on the preservation of
  conditionally periodic motions under a small perturbation of the
  Hamiltonian}.
\newblock {\em Russ. Math. Surv.}, 18:9, 1963.

\bibitem{Nekhoroshev:1971aa}
N.~Nekhoroshev.
\newblock {Behavior of Hamiltonian systems close to integrable}.
\newblock In {\em Functional Analysis and Its Applications}, volume~5, page
  338. Kluwer Academic Publishers-Plenum Publishers, 1971.

\bibitem{Nekhoroshev:1977aa}
N.~Nekhoroshev.
\newblock {An exponential estimate of the time of stability of
  nearly-integrable Hamiltonian systems}.
\newblock {\em Russ. Math. Surv.}, 32(6):1, 12 1977.

\bibitem{Bazzani:1990aa}
A.~Bazzani, S.~Marmi, and G.~Turchetti.
\newblock {Nekhoroshev estimate for isochronous non resonant symplectic maps}.
\newblock {\em Cel. Mech.}, 47(4):333, 1990.

\bibitem{Turchetti:1990aa}
G.~Turchetti.
\newblock {Nekhoroshev Stability Estimates for Symplectic Maps and Physical
  Applications}.
\newblock In J.-M. Luck, P.~Moussa, and M.~Waldschmidt, editors, {\em Number
  Theory and Physics}, volume~47 of {\em Springer Proceedings in Physics},
  pages 223--234, Berlin, Heidelberg, 1990. Springer Berlin Heidelberg.

\bibitem{da_and_losses}
M.~Giovannozzi.
\newblock {A proposed scaling law for intensity evolution in hadron storage
  rings based on dynamic aperture variation with time}.
\newblock {\em Phys. Rev. ST Accel. Beams}, 15:024001, 2 2012.

\bibitem{Corless1996}
R.~M. Corless, G.~H. Gonnet, D.~E.~G. Hare, D.~J. Jeffrey, and D.~E. Knuth.
\newblock On the lambertw function.
\newblock {\em Advances in Computational Mathematics}, 5(1):329--359, 12 1996.

\bibitem{10.5555/1403886}
W.~H. Press, S.~A. Teukolsky, W.~T. Vetterling, and B.~P. Flannery.
\newblock {\em {Numerical Recipes 3rd Edition: The Art of Scientific
  Computing}}.
\newblock Cambridge University Press, USA, 3 edition, 2007.

\bibitem{bazzani2020diffusion}
A.~Bazzani, M.~Giovannozzi, and E.~H. Maclean.
\newblock {Analysis of the non-linear beam dynamics at top energy for the CERN
  Large Hadron Collider by means of a diffusion model}.
\newblock {\em Eur. Phys. J. Plus}, 135(1):77, 2020.

\bibitem{our_paper9}
C.~E. Montanari, A.~Bazzani, and M.~Giovannozzi.
\newblock {Probing the diffusive behaviour of beam-halo dynamics in circular
  accelerators}.
\newblock {\em Eur. Phys. J. Plus}, 137(11):1264, 2022.

\bibitem{Montanari:2728138}
C.E. Montanari.
\newblock {Diffusive approach for non-linear beam dynamics in a circular
  accelerator}.
\newblock Master thesis, Department of Physics and Astronomy, Alma Mater
  Studiorum -- University of Bologna, 2019.
\newblock CERN-THESIS-2019-383.

\bibitem{montanari:ipac22-mopost043}
C.~E. Montanari, A.~Bazzani, M.~Giovannozzi, A.~A. Gorzawski, and S.~Redaelli.
\newblock {Testing the Global Diffusive Behaviour of Beam-Halo Dynamics at the
  CERN LHC Using Collimator Scans}.
\newblock In {\em Proc. IPAC'22}, number~13 in International Particle
  Accelerator Conference, pages 172--175. JACoW Publishing, Geneva,
  Switzerland, 7 2022.

\bibitem{montanari:2025}
C.~E. Montanari, R.~B. Appleby, A.~Bazzani, M.~Giovannozzi, P.~Hermes,
  A.~Poyet, S.~Redaelli, and G.~Sterbini.
\newblock {Measurement of the nonlinear diffusion of the proton beam halo at
  the CERN LHC}.
\newblock Accepted for publication in Eur. Phys. J. Plus, 2025.

\bibitem{Milani1992}
A.~Milani and A.~M. Nobili.
\newblock An example of stable chaos in the solar system.
\newblock {\em Nature}, 357(6379):569--571, 1992.

\bibitem{MILANI199713}
A.~Milani, A.~M. Nobili, and Z.~Knežević.
\newblock Stable chaos in the asteroid belt.
\newblock {\em Icarus}, 125(1):13--31, 1997.

\bibitem{sym15061195}
S.~Elaskar and E.~del Río.
\newblock Review of chaotic intermittency.
\newblock {\em Symmetry}, 15(6), 2023.

\bibitem{skokos2016chaos}
C.~H. Skokos, G.~A Gottwald, and J.~Laskar.
\newblock {\em Chaos detection and predictability}, volume~1.
\newblock Springer, 2016.

\bibitem{Froeschle1997}
C.~{Froeschl{\'e}}, R.~{Gonczi}, and E.~{Lega}.
\newblock {The fast Lyapunov indicator: a simple tool to detect weak chaos.
  Application to the structure of the main asteroidal belt}.
\newblock {\em Planetary and Space Science}, 45(7):881--886, 1997.
\newblock Asteroids, Comets, Meteors 1996 - II.

\bibitem{Barrio2016}
R.~Barrio.
\newblock {Theory and Applications of the Orthogonal Fast Lyapunov Indicator
  (OFLI and OFLI2) Methods}.
\newblock {\em Chaos Detection and Predictability}, 915:55--92, 3 2016.

\bibitem{Panichi2018}
{F. Panichi and G. Turchetti}.
\newblock {Lyapunov and reversibility errors for Hamiltonian flows}.
\newblock {\em Chaos Solitons and Fractals}, 112:83--91, 7 2018.

\bibitem{Panichi19}
G.~Turchetti and F.~Panichi.
\newblock {Fast Indicators for Orbital Stability: A Survey on Lyapunov and
  Reversibility Errors}.
\newblock In C.~G. Buzea, M.~Agop, and L.~Butler, editors, {\em Progress in
  Relativity}, chapter~10. IntechOpen, Rijeka, 2019.

\bibitem{8766229}
{IEEE Standard for Floating-Point Arithmetic}.
\newblock {\em IEEE Std 754-2019 (Revision of IEEE 754-2008)}, pages 1--84,
  2019.

\bibitem{Poschel1982}
J.~P{\"o}schel.
\newblock {The concept of integrability on cantor sets for Hamiltonian
  systems}.
\newblock {\em Celestial Mech. Dyn. Astron.}, 28(1):133--139, 9 1982.

\bibitem{laskar1992measure}
J.~Laskar, C.~Froeschl{\'e}, and A.~Celletti.
\newblock {The measure of chaos by the numerical analysis of the fundamental
  frequencies. Application to the standard mapping}.
\newblock {\em Physica D}, 56(2-3):253--269, 05 1992.

\bibitem{bartolini1998computer}
R.~Bartolini and F.~Schmidt.
\newblock {A computer code for frequency analysis of non-linear betatron
  motion}.
\newblock Technical report, CERN-SL-Note-98-017-AP, 1998.

\end{thebibliography}
%
\end{document}